# A Human-Centered Privacy Approach (HCP) to AI

*Corresponding author: Luyi Sun, Department of Psychology and Behavioral Sciences, Zhejiang University, Hangzhou, China, luyis@zju.edu.cn , 0000-0001-8997-9152*

*Wei Xu, Department of Psychology and Behavioral Sciences, Zhejiang University, Hangzhou, China, weixu6@yahoo.com*

*Zaifeng Gao, Department of Psychology and Behavioral Sciences, Zhejiang University, Hangzhou, zaifengg@zju.edu.cn*

## Abstract

*As the paradigm of Human-Centered AI (HCAI) gains prominence, its benefits to society are accompanied by significant ethical concerns, one of which is the protection of individual privacy. This chapter provides a comprehensive overview of privacy within HCAI, proposing a human-centered privacy (HCP) framework, providing integrated solution from technology, ethics, and human factors perspectives. The chapter begins by mapping privacy risks across each stage of AI development lifecycle, from data collection to deployment and reuse, highlighting the impact of privacy risks on the entire system. The chapter then introduces privacy-preserving techniques such as federated learning and differential privacy. Subsequent chapters integrate the crucial user perspective by examining mental models, alongside the evolving regulatory and ethical landscapes as well as privacy governance. Next, advice on design guidelines is provided based on the human-centered privacy framework. After that, we introduce practical case studies across diverse fields. Finally, the chapter discusses persistent open challenges and future research directions, concluding that a multidisciplinary approach, merging technical, design, policy, and ethical expertise, is essential to successfully embed privacy into the core of HCAI, thereby ensuring these technologies advance in a manner that respects and ensures human autonomy, trust and dignity.*

## Keywords

**1. Introduction**

1.1 The Centrality of Privacy in the AI Era

Privacy stands as a fundamental human right and a cornerstone of democratic society, yet its preservation faces unprecedented challenges in the age of artificial intelligence (AI). Broadly known, privacy's multidimensional nature often encompasses informational privacy (concerning data collection, storage, sharing, and ownership), psychological privacy (mental privacy), and physical privacy (including location, surveillance, biometric data, and personal territory) (Smith et al., 1996; Solove, 2006). In the digital age, privacy-enhancing techniques have emerged as an essential dimension for protecting these various privacy facets (Tavani, 2007).

The rapid advancement of AI technology has intensified public concerns regarding digital surveillance and privacy erosion. The proliferation of cybercrime, mass surveillance, internet censorship, and espionage creates what Hagen and Lysne (2016) describe as a surveillance-privacy paradox, wherein policymakers must balance security needs with individual freedoms. From a psychological perspective, surveillance technologies demonstrably impact individuals' sense of privacy, producing a "chilling effect" whereby awareness of observation fundamentally alters behavior patterns with far-reaching social implications (Wright & Raab, 2012).

Within socio-technical systems, which are characterized by interconnected relationships between humans and machines, privacy implementation presents unique challenges (Kolesnyk et al., 2025). Privacy spans legal, psychological, and physical domains, with technical definitions often failing to capture its full complexity within these systems (Knijnenburg et al., 2022). This multifaceted nature necessitates addressing privacy from both social and technical perspectives, establishing it as a critical consideration that permeates every aspect of AI system design and deployment.

Therefore, safeguarding privacy in the AI era requires a holistic approach that moves beyond a purely technical focus. It demands a supportive ecosystem of legal regulations and organizational governance to establish clear boundaries and accountability. Foundational regulatory instruments, such as the European Union's General Data Protection Regulation (GDPR；European Union, 2016) and the California Consumer Privacy Act (CCPA；California Legislature, 2018), have established critical principles for data rights and corporate responsibilities. Complementing these laws, governance frameworks like the National Institute of Standards and Technology (NIST, 2020) privacy framework offer structured methodologies for organizations to manage privacy risks proactively and embed privacy-by-design into their operations. The integration of robust technology, binding law, and strategic governance thus forms the essential triad for effective and human-centered privacy protection.

The emergence of Human-Centered AI (HCAI) represents a paradigm shift in AI development, moving from technology-centric approaches to a complementary design philosophy and methodology. This approach prioritizes human needs, values, ethics, experiences, and control throughout the AI lifecycle, with the goal of creating systems that augment and enhance human capabilities, rather than replacing or harming them (Riedl, 2019; Xu et al., 2023). Figure 1 shows the Technology-Human Factors-Ethics (THE) triangle framework proposed by Xu (2019), which integrates key perspectives: ethical design, technology, and human factors (Xu et al., 2024). By fostering self-efficacy, encouraging creativity, clarifying user responsibilities, and promoting social participation, HCAI establishes a foundation for more ethical and effective human-AI interaction (Shneiderman, 2020).

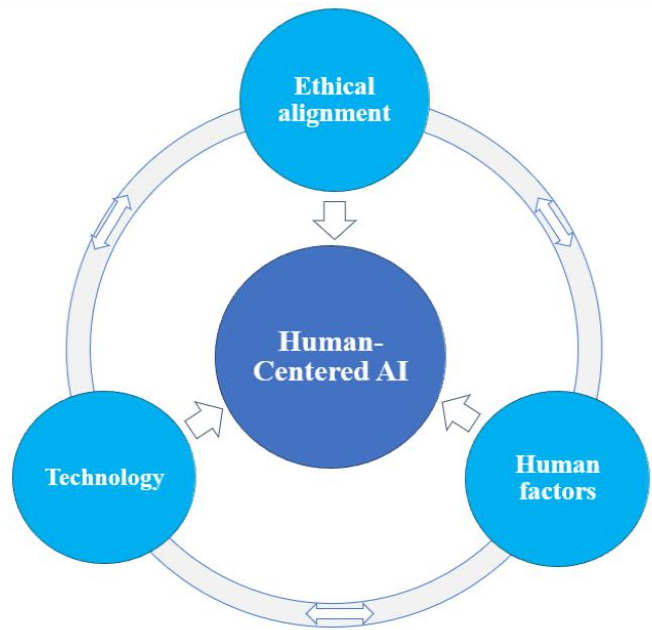

Fig 1. The "Technology-Human Factors-Ethics" framework (Xu, 2019)

1.2 Human-Centered Privacy Framework

This chapter adopts a comprehensive approach to privacy, highlighting its critical importance through a human-centered lens. Building on the framework proposed by Xu (2019), this chapter provides an in-depth examination of this framework from a privacy perspective. Figure 2 illustrates the interconnections between privacy and the key components of HCAI systems: ethics, technology, and human factors, while also demonstrating how this core framework is enveloped by three crucial protective layers, which are design guidelines, privacy governance, and privacy regulations, that function like concentric defensive barriers safeguarding the entire system.

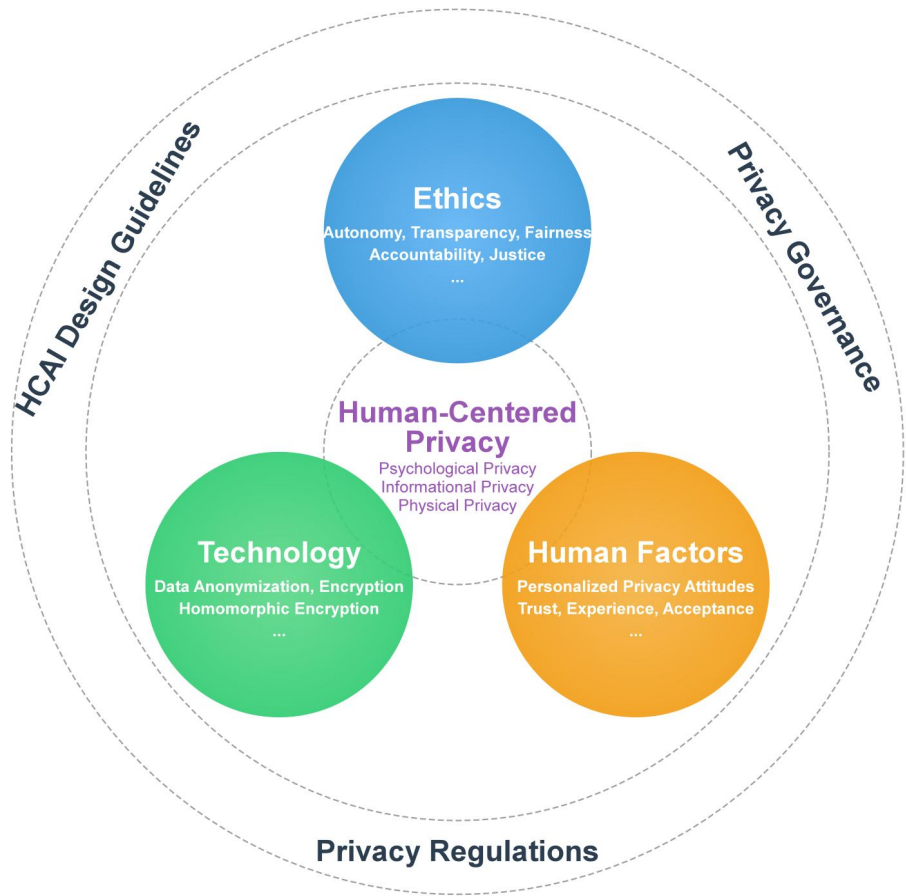

Fig 2. Human-Centered Privacy (HCP) Framework

A human-centered privacy framework for AI is established through the systematic integration of three interdependent pillars: the ethical elements explicitly outlined in Table 1 (Section 3), the privacy-enhancing technologies listed in Table 2 (Section 4), and the human factors detailed in Table 3 (Section 5). This integrated approach posits that ethics provides the foundational "why," supplying the normative justification by positioning privacy as a cornerstone that enables other principles like trust, fairness, and human dignity. These values then guide the selection and application of technological tools, which enact the "what" by providing tangible protections, while navigating inherent trade-offs between privacy, utility, and performance. Crucially, the effectiveness of both ethical issues and technical solutions is ultimately mediated by human factors, which define the "how" of user experience. It is through aligning with user mental models, expectations, and cultural contexts that technical guarantees become comprehensible and ethical principles become meaningful, thereby fostering genuine trust and user acceptance. A truly robust and trustworthy privacy solution, therefore, only emerges from this recursive relationship, where each dimension continuously informs and reinforces the others.

Surrounding this core framework, three protective layers create a comprehensive defense system. Design guidelines ensure privacy considerations are embedded from the outset in system development, establishing proactive measures that prevent privacy violations before they occur. Privacy governance provides the organizational structures, processes, and accountability mechanisms that oversee privacy implementation throughout the system lifecycle. Privacy regulations form the outermost protective layer, establishing the legal requirements and compliance standards that mandate privacy protection.

The framework's privacy perspective is notably comprehensive, integrating the core dimensions of privacy: informational, physical, and psychological. This integrated view is essential, as users' privacy attitudes are not formed in a vacuum. Rather, they are context-specific and frequently constrained by physical boundaries. Furthermore, these attitudes, which include but are not limited to preferences for controlling biometrics data and other personal information, are themselves shaped by the user's internal mental state, a core aspect of psychological privacy.

Based on this framework, the rest of the chapter provides a systematic approach to privacy in HCAI, addressing the following key objectives:

Establishing privacy as a core ethical principle and outlining relevant global regulations and frameworks.

(2) Identifying privacy risks throughout the AI lifecycle and examining technological solutions with their limitations.

(3) Understanding users' perceptions, mental models, and attitudes toward AI privacy.

(4) Developing robust privacy governance frameworks and actionable design guidelines.

(5) Synthesizing insights from case studies and proposing future research directions and challenges.

By placing privacy at the center of our analysis, this chapter demonstrates that privacy is not merely a regulatory compliance issue or technical challenge, but rather the essential foundation upon which ethical and human-centered AI must be built.

1.3 Privacy Theories from Technology, Human Factors, Ethics Perspectives

Based on the human-centered privacy framework, privacy theories can also be usefully categorized into three interconnected perspectives: technological, psychological, and ethical. From a technological standpoint, the focus shifts from static defenses to adaptive, lifecycle-aware solutions. The personalized differential privacy framework allows users to set individualized privacy levels, balancing protection and utility (Liu et al., 2024). Concurrently, the concept of privacy beyond memorization identifies risks in Large Language Models that extend throughout the model lifecycle, from inference-time exposure to agentic misuse (Staab et al., 2023). Complementing these, the activity theory of privacy reconceptualizes privacy as a dynamic, context-dependent activity, urging the design of tools that empower users to manage their privacy fluidly (Clemmensen, 2021).

From a user psychology perspective, privacy behaviors are shaped by tensions between desires and perceived risks. The privacy paradox captures the conflict between users' desire for personalization and their discomfort with data exposure (Aguirre et al., 2016). Privacy self-efficacy underscores that an individual's belief in their ability to protect their data is a key predictor of their privacy behaviors (Chen & Chen, 2015). Furthermore, behavioral fingerprinting demonstrates that even de-identified behavioral traces can be used for re-identification, challenging users' sense of anonymity (Oliveira, 2025). These insights collectively highlight that effective privacy design must address psychological realities to reduce cognitive burden and reinforce autonomy.

From the ethical perspective, theories move beyond compliance to address the moral foundations of data governance. The privacy as control paradigm asserts privacy as an individual's right to manage their information, reflecting respect for autonomy (Allen, 1999). As data ecosystems grow complex, the privacy as trust paradigm reframes protection as a systemic duty of data processors, grounded in accountability and responsibility (Bella et al., 2008). The reasonable expectation of privacy further integrates societal values by balancing individual subjectivity with collective standards of fairness (Katz v. United States, 1967). Together, these perspectives argue that privacy is both a personal right and a social good, essential for institutional integrity.

1.4 Scope and Objectives

This comprehensive exploration of AI privacy through a human-centered perspective follows a structured progression. Section 2 establishes the foundation by examining the AI lifecycle, identifying specific stages where privacy risks emerge and demonstrating their propagation throughout system development and deployment. Section 3 discusses ethical elements, clarifying the relation between privacy and other elements. Section 4 addresses technological countermeasures, exploring privacy-preserving techniques including differential privacy, federated learning, and homomorphic encryption.

Recognizing technology's limitations, Section 5 investigates the human element, examining user perspectives, privacy concerns, and mental models in AI interaction. Section 6 contextualizes these findings within the regulatory and ethical ecosystem, analyzing frameworks such as General Data Protection Regulations (GDPR) and AI ethics principles. Section 7 introduces adaptive privacy governance mechanisms to manage privacy risks effectively. And Section 8 synthesizes technological, human, and ethical insights into actionable design guidelines for HCAI privacy implementation.

Practical applications are demonstrated in Section 9 through case studies across healthcare, finance, education, and smart cities. Section 10 discusses open challenges and future research directions, while Section 11 concludes by summarizing key contributions and reinforcing the imperative of embedding privacy into AI systems' fundamental architecture. This structured approach provides both theoretical grounding and practical guidance for implementing privacy as the cornerstone of human-centered AI systems.

## 2. Privacy Risks Across the AI Lifecycle

Section 2 presents an overview of the entire AI lifecycle, highlighting the privacy risks at each phase. The AI lifecycle begins with data collection and processing, followed by model training, inference and deployment, and concludes with feedback loops and data reuse. The privacy risks across different phases are discussed respectively.

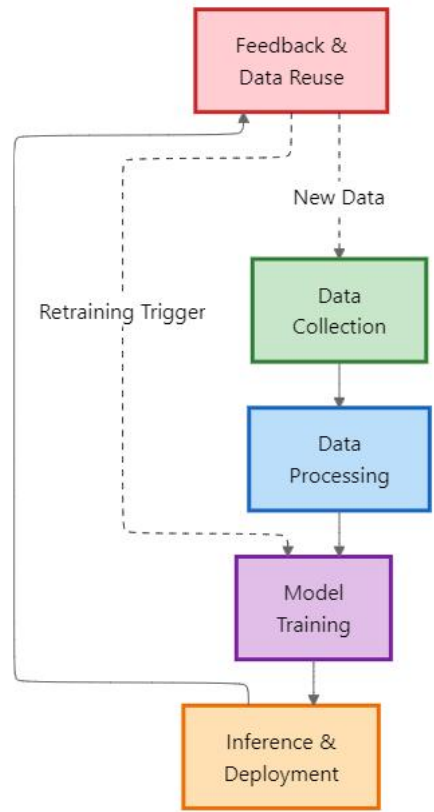

Fig 3. AI Lifecycle

2.1 Data Collection Phase

Once the project's objective has been identified and its scope defined, the data collection phase begins. This stage provides the information that data scientists need to implement use cases and build the AI model (Haakman et al., 2021). It is also the first point in the AI lifecycle where privacy risks can emerge. In this phase, the collected data may take various forms, such as numerical, categorical, time series, or textual (Luetto et al., 2025). Each data type has its own characteristics and complexities which can influence susceptibility to vulnerabilities.

The data collection phase must ensure that data collection does not violate privacy and ethical guidelines. This phase requires informed consent from users and must adhere to key ethical and legal principles such as transparency, data minimization, purpose limitation, and contextual integrity. Researchers should be clear and open about what data is being collected, why it is needed, how it will be used, and who will have access to it. Explicit permission from individuals must be obtained before collecting their data. To protect privacy, only data strictly necessary should be gathered, and unrelated information should be avoided. Furthermore, sharing of information must align with the social norms of the specific context in which it was originally provided. (Burton et al., 2024)

2.2 Data Processing Phase

The data processing phase involves several key activities, including data cleaning, transformation, integration, anonymization and pseudonymization, feature selection, and validation. Consistent with the definition of this phase, Priyank Jain et al. (2016)

note that it also encompasses Privacy-Preserving Data Publishing (PPDP), which releases useful versions of datasets for analysis and research while protecting the sensitive personal information of the individuals within that data, and the extraction of knowledge from the data.

During this phase, AI technologies introduce new risks related to re-identification and aggregation. For example, matching anonymous or pseudonymous data can be linked back to an individual's identity using auxiliary information (Morehouse et al., 2024). In addition, AI can repurpose foundational models, train on datasets containing content acquired without consent, and lead to profiling risks, and create new security vulnerabilities as a result of the technology's application. Finally, by operating on low-quality data or forecasting future events, AI may exacerbate risks of secondary use, data leakage, exclusion, and insecurity (Lee et al., 2024).

2.3 Model Training Phase

In the model training phase, factors such as limited sample size, poor handling of missing data, and inadequate evaluation of overfitting significantly contribute to the overall high risk of bias (Navarro et al., 2021), thereby undermining system performance. Malicious actors may also inject biased or erroneous data to manipulate model behavior, leading to bias amplification. Moreover, the inclusion of sensitive attributes during training can introduce privacy risks and intensify bias, as seen in discriminatory hiring practices in the job market (Rogen et al., 2020).

Barocas et al. (2017) offer a useful framework for understanding how these consequences manifest, distinguishing between allocative harms, when opportunities or resources are withheld from certain individuals or groups, and representational harms, when individuals or groups are stigmatized or stereotyped. During this phase, harms or negative consequences caused by such systems can be classified into different types, including historical bias, representation bias, measurement bias, aggregation bias, learning bias, evaluation bias, deployment bias (Suresh & Guttag, 2021).

2.4 Inference and Deployment Phase

Privacy risks in the AI inference and deployment phase can be categorized in several ways. According to a survey of privacy attacks in machine learning, the important attack types during model training and deployment include membership inference attacks, reconstruction attacks, property inference attacks, and model extraction attacks (Rigaki & Garcia, 2023; Khalid et al., 2023). Membership inference attacks seek to determine whether a specific input sample was part of a model's training set. Reconstruction attacks aim to recreate training samples and/or their labels. Property inference attacks attempt to extract information that the model has unintentionally learned, which is unrelated to the primary training task. Model extraction attacks occur when an adversary, with access to a model's inputs and prediction outputs, treats input-output pairs as a system of equations in which the unknowns are the model's parameters or the hyperparameters of the objective function (Wang & Gong, 2018; Yan et al., 2023). These enable the adversary to replicate the model's functionality, steal proprietary parameters, and steal ML models through prediction APIs.

Additionally, privacy risks during this phase may also involve unintended consequences. Unlike property inference attacks, these risks typically arise from accidental flaws in AI deployment rather than from malicious actions (Cheatham et al., 2019; Wang et al., 2025).

2.5 Feedback Loops and Data Reuse Phase

AI systems often rely on feedback loops, where model outputs influence future training data, and data reuse, where the same dataset is used across multiple models or iterations. It is the last phase of the AI lifecycle. Privacy risks emerge in this phase include data leakage, long-term privacy erosion, and transparency challenges. Data leakage occurs when sensitive information is inadvertently exposed through AI outputs, model weights, or reused datasets. In feedback loops, models may memorize and reproduce private data, while data reuse increases re-identification risks (Carlini et al., 2021).

In the feedback loop, reusing data across multiple AI iterations leads to privacy decay, where anonymization weakens over time (often referred to as long-term privacy erosion). According to Dwork et al. (2014), each interaction with an AI system may leak small amounts of data, aggregating into a privacy risk. Also, combining multiple model outputs can deanonymize users, leading to linkage attacks (Powar & Beresford, 2023).

In this phase, there will also be transparency challenges caused by opaqueness in how data is processed, making it difficult to track where the data is used across multiple models (Holland et al., 2020). According to consent management, users may not expect their data to be reused under evolving AI feedback loops, which means there will be fuzzy domains in user consent.

3. Ethics Perspective

This section examines privacy as a foundational ethical pillar within Human-Centered AI, emphasizing its dual role as both an independent moral value and an essential enabler of other ethical principles. Table 1 provides a systematic overview of ethical elements and their interrelationships with privacy, which serves as the conceptual foundation for the subsequent discussion in this section.

**Table 1 Overview of Ethical Elements**

| Category | Ethical Elements | Core Concepts & Relationship to Privacy |
|---|---|---|
| **1. The fundamental principles that privacy directly embodies or protects** | Privacy (e.g., Malik et al., 2025) | Serves as a foundational ethical pillar in HCAI. It is both an independent moral value and an essential enabler of other ethical principles. Its protection is fundamentally grounded in the preservation of human dignity. |
| | Human Dignity (e.g., Usmani et al., 2023) | Has an inviolable link with privacy. Privacy protection must be fundamentally grounded in its preservation. AI technologies without proper safeguards risk fundamentally undermining privacy rights and human dignity. |
| | Autonomy (e.g., Allen, 1999) | Can be compromised by privacy violations. Protecting privacy ensures that individuals can make decisions without their personal data being misused, thereby safeguarding their personal autonomy and self-determination. |

| Category | Ethical Elements | Core Concepts & Relationship to Privacy |
|---|---|---|
| 2. The processes and measures through which privacy is implemented and upheld | Transparency (e.g., Bingley et al., 2023) | Requires making privacy-relevant system behaviors understandable to users. The pursuit of algorithmic transparency can compromise privacy, thus demanding privacy-preserving explanation methods. It is crucial for informed consent. |
| | Accountability (e.g., Shneiderman, 2020) | Ensures that there is responsibility for system actions and decisions, including those related to privacy management. |
| | Security (e.g., Jobin et al., 2019) | Provides the technical foundation for implementing privacy protections (e.g., through Privacy-Enhancing Technologies). It is directly threatened by privacy violations, which can expose personal information. |
| | Reliability (e.g., Mehrabi et al., 2021) | A reliable AI system performs consistently and predictably. Privacy underpins reliability by ensuring that the personal data used for training and operation is accurate, used appropriately, and protected from corruption or misuse that could lead to erratic or erroneous system behavior. |
| 3. The broader ethical goals that are enabled by a well-functioning privacy framework | Fairness (e.g., Agrawal, 2024) | Can be undermined by privacy violations, particularly through discriminatory data use. Effective bias mitigation is intrinsically linked to privacy, as biases are often propagated through the uninformed or inappropriate use of personal data. |
| | Justice (e.g., Agrawal, 2024) | As a broader ethical principle, it is part of the constellation of considerations that privacy supports and interacts with. A system that respects privacy helps prevent injustices stemming from the misuse of personal data. |
| | Trust (e.g., Mohseni et al., 2021) | Privacy acts as a critical enabler for trust. Without credible privacy assurances, users cannot develop the trust necessary for meaningful engagement with AI. |

3.1 Privacy as a Foundational Ethical Element

Contemporary HCAI development grapples with a constellation of ethical considerations (Bingley et al., 2023; Shneiderman, 2020). While each ethical dimension carries significant importance, this work places particular emphasis on privacy as it fundamentally underpins many other ethical considerations, such as human dignity and autonomy. Privacy violations can compromise autonomy and underpin human dignity. Thus, while maintaining a holistic view of AI ethics, privacy deserves special attention as both a standalone concern and an enabler of other ethical principles.

Jobin et al. (2019) revealed that developers prioritize ethics (20%), privacy (11%), and security (6%) in AI design, yet these factors often remain peripheral to actual user experiences. This gap between intention and implementation is particularly concerning for privacy, as AI technologies without proper safeguards risk fundamentally undermining privacy rights and human dignity (Usmani et al., 2023). The challenge intensifies as users demand increasingly intelligent and adaptabile AI systems while simultaneously expecting robust privacy protections, a tension that requires careful navigation through human-centered design approaches.

3.2 Privacy Implementation and Safeguarding Processes

The implementation of privacy in AI systems relies on processes and measures such as transparency, accountability, security, and reliability. Transparency requires that privacy-relevant system behaviors be made understandable to users, yet the pursuit of algorithmic transparency may conflict with privacy, necessitating privacy-preserving explanation methods (Bingley et al., 2023). Accountability ensures responsibility for system decisions, including those affecting privacy (Shneiderman, 2020). Security provides the technical foundation for privacy through Privacy-Enhancing Technologies (PETs), while reliability is supported by privacy in ensuring that training and operational data remain accurate and protected from misuse (Jobin et al., 2019; Mehrabi et al., 2021).

In explainable AI (XAI), privacy awareness is not merely a complementary feature but a critical design goal deeply interwoven with algorithmic transparency, user trust, and bias mitigation (Mohseni et al., 2021). The pursuit of algorithmic transparency can be compromised if model explanations reveal sensitive training data, thus demanding privacy-preserving explanation methods. Similarly, user trust and reliance hinge upon credible privacy assurances. Therefore, implementing PETs is essential to safeguard these interconnected objectives.

3.3 Broader Ethical Goals Enabled by Privacy

A well-functioning privacy framework enables the realization of broader ethical goals such as fairness, justice, and trust. Privacy violations can undermine fairness, particularly through discriminatory data use (Agrawal, 2024). Similarly, justice as an ethical principle is supported by privacy, which helps prevent injustices arising from misuse of personal data. Trust, in particular, is critically enabled by privacy. Without credible privacy assurances, users cannot develop the trust necessary for meaningful engagement with AI systems (Mohseni et al., 2021; Schoenherr et al., 2023).

3.4 Ethics Approaches

To translate ethical principles into actionable strategies for privacy protection in HCAI, several foundational approaches are critical. These methodologies bridge abstract values and system design, ensuring privacy is proactively upheld.

First, privacy as a fundamental right provides the philosophical foundation. This deontological perspective, underpinning regulations, asserts that privacy is an inherent human right that must be respected as a duty, not merely balanced against utility (Okpo & Joseph, 2025). Second, implementation frameworks translate principles into practice. Privacy by Design (PbD) mandates embedding privacy into the system architecture by default, from the earliest stages of development (Alkhariji et al., 2021). Complementing this, ethical transparency requires making privacy-relevant system behaviors (e.g., data collection, inference, retention) understandable to users and regulators. Transparency is crucial for fostering informed consent and trust, as it enables individuals to comprehend the trade-offs between personalization and privacy (Radanliev, 2025). Furthermore, contextual integrity guides designers to ensure information flows align with context-specific social norms, respecting differing expectations across domains (Kumar et al., 2024). Finally, participatory privacy ensures these approaches remain grounded in human values by engaging diverse stakeholders in defining privacy norms (Wacnik et al., 2025). This process, supported by continuous ethical oversight, upholds human dignity and fosters trust in AI systems.

Together, these approaches form a coherent ethical methodology that moves beyond compliance, positioning privacy as a non-negotiable, context-aware value in human-centered AI systems.

## 4. Technology Perspective

This section provides a thorough introduction to different privacy-enhancing technologies, including the core concept of each technique as well as their advantages and disadvantages. Table 2 establishes the conceptual framework for each technique, while the subsequent subsections delve deeper into their practical implementations, trade-offs, and specific applications. The exploration covers the balance of these techniques regarding privacy assurance, computational overhead, communication costs, and utility, forming a foundational understanding for selecting appropriate privacy-preserving solutions.

**Table 2 Overview of Privacy-Enhancing Technologies**

| Category | Technique | Core Concepts |
|---|---|---|
| **1. Data-Level Protection** | Data anonymization (e.g., Olatunji et al., 2024) | Processes data to remove or obscure personally identifiable information (PII), aiming to prevent the identification of individuals from datasets. |
| | Data encryption (e.g., Atadoga et al., 2024) | Protects data by converting it into a cipher-text that is unreadable without a secret key. It secures data both at rest (in storage) and in transit (during transmission). |
| **2. Process-Level Protection** | Differential privacy (e.g., Dodiya et al., 2024) | A mathematical framework that provides robust, quantifiable privacy guarantees. It works by injecting carefully calibrated noise into the outputs of data analysis or model training. |
| | Federated learning (e.g., Zhu et al., 2019) | A decentralized machine learning approach. Instead of sending raw data to a central server, models are trained locally on user devices (e.g., phones, hospital servers), and only the model updates (e.g., gradients) are shared and aggregated. |
| **3. Encrypted Computation** | Homomorphic encryption (e.g., Ghanem & Moursy, 2019) | Allows computations to be performed directly on encrypted data without needing to decrypt it first. The result of the computation remains encrypted and can only be read by the owner of the decryption key. |
| | Secure multi-party computation (e.g., Jiang et al., 2022) | Enables multiple parties, each holding their own private data, to jointly compute a function (e.g., train a model) without any party having to reveal its private input data to the others. |

4.1 Data-Level Protection

To address privacy threats such as model inversion attacks, membership inference attacks, and unintended data memorization that can originate from malicious external actors, unauthorized internal access, or inadvertent data leaks during training and deployment (Rigaki & Garcia, 2023), data-level protection techniques form the first line of defense. These techniques focus on securing the data itself before any processing occurs.

Data anonymization processes data to remove or obscure personally identifiable information (PII), aiming to prevent the identification of individuals from datasets (Olatunji et al., 2024). While straightforward to implement, it faces challenges in maintaining data utility while ensuring true anonymity, as sophisticated re-identification attacks can sometimes reverse the anonymization process.

Data encryption protects data by converting it into cipher-text that is unreadable without a secret key, securing data both at rest (in storage) and in transit (during transmission) (Atadoga et al., 2024). This foundational method ensures confidentiality but requires careful key management and introduces computational overhead during encryption and decryption operations.

These data-centric approaches are particularly valuable in scenarios where raw data must be shared or stored in potentially insecure environments, though they primarily address external threats rather than risks arising during data processing itself.

4.2 Process-Level Protection

Process-level protection techniques focus on safeguarding privacy during data analysis and model training, addressing limitations of static data protection methods.

Differential privacy has emerged as a widely adopted framework for achieving robust privacy preservation in data analysis. By injecting carefully calibrated noise into the outputs of data analysis or model training, it provides mathematically rigorous, quantifiable privacy guarantees (Dodiya et al., 2024). It is particularly valuable in high-stakes domains like healthcare and finance (Yang et al., 2024). The fundamental strength lies in its resilience against adversarial attacks, as privacy guarantees do not rely on the adversary's ignorance of the mechanism itself (Dwork & Roth, 2014). However, this approach inherently involves a privacy-utility tradeoff, where increasing noise enhances privacy protection but reduces accuracy. Key challenges include vulnerability to composition attacks and the need to carefully manage a finite privacy budget across all analyses (Dwork & Roth, 2014). To address these limitations, advanced noise-reduction techniques dynamically adjust noise levels according to data sensitivity and model requirements (Alzoubi et al., 2025; Jiang et al., 2021). Additionally, post-processing and perturbation techniques can refine noise outputs, minimizing unnecessary distortion while preserving privacy (Ghazi et al., 2021). Common techniques include Laplace and Gaussian noise addition, designed to optimize privacy guarantees without compromising model utility (Pan et al., 2024).

Federated learning offers a decentralized machine learning approach that enables model training across multiple edge devices and servers without centralizing raw data (Zhu et al., 2019). Instead of transferring sensitive data to a central server, each device computes model updates locally and transmits only these updates for aggregation. This framework is particularly valuable in privacy-sensitive domains such as healthcare and finance, where sharing raw data across organizations might cause significant risks (Mammen, 2021). The architecture of local data control also enables model personalization without privacy trade-offs, as data never leaves the device. However, while raw data remains local, transmitted model updates (e.g., gradients) can still be vulnerable to inference attacks, potentially allowing a malicious server to reconstruct sensitive training data or identify membership in the dataset (Zhu et al., 2019).

4.3 Encrypted Computation

Encrypted computation techniques represent the most advanced approach to privacy preservation, enabling secure data processing while maintaining cryptographic protection throughout computations.

Homomorphic encryption allows computations to be performed directly on encrypted data without requiring decryption (Ghanem & Moursy, 2019). This makes it valuable for privacy-preserving cloud computing and big data analytics, where sensitive data must remain confidential during processing (Akl & Assem, 2020). Critically, it enables a single party to maintain control over encrypted data while outsourcing computations to untrusted servers (Liang et al., 2025). However, the core disadvantage is substantial computational overhead and latency, often rendering it impractical for real-time applications (Naehrig et al., 2011).

Secure multi-party computation enables multiple parties, each holding private data, to jointly compute a function without revealing their individual inputs to others (Jiang et al., 2022). This approach operates in a trustless environment, eliminating the need for participants to share raw data or rely on a central authority. The primary disadvantages, however, are high communication costs and scalability challenges. The need for continuous interaction between parties creates significant network bottlenecks, especially over wide-area networks with many participants (Evans et al., 2018).

These encrypted computation methods provide the strongest privacy guarantees but come with significant performance costs, making them suitable for applications where privacy requirements justify the additional computational and communication overhead. Several privacy-preserving techniques including homomorphic encryption and secure multi-party computation have been proposed to mitigate privacy risks while enabling data utility (Aziz et al., 2023).

5. **Human Factors Perspective**

This section shifts the focus from technical mechanisms to the human aspects of AI privacy, examining how users perceive and understand privacy within AI systems. Table 3 systematically organizes the psychological and behavioral elements, from mental models and cultural influences to trust formation and user experience, that collectively shape how individuals perceive and interact with privacy-preserving AI systems. This comprehensive taxonomy provides the foundation for the detailed analysis in the following subsections.

**Table 3 Overview of Human Factors**

| Category | Human Factors | Explanation & Relationship to Other Factors |
|---|---|---|
| **1. Cognitive Foundations** | User Mental Model (e.g., Rudolph et al., 2018) | The user's internal understanding of how an AI system uses their data. This model is often incomplete or inaccurate, leading users to agree to terms without full comprehension. It is the foundation upon which all other |

| Category | Human Factors | Explanation & Relationship to Other Factors |
|---|---|---|
| | | perceptions (trust, expectations, acceptance) are built. |
| | Cultural Differences (e.g., Krasnova & Veltri, 2010) | Shapes user mental models. User mental models are not universal. Instead, they are significantly influenced by cultural dimensions like individualism-collectivism, nationality, and religion. This means a single privacy design will not align with all users' mental models. |
| | Awareness (e.g., Wongso et al., 2024) | The level of user understanding regarding privacy risks and AI functionalities. It is an antidote to poor mental models and human error. Enhancing awareness through training and clear feedback is crucial for enabling informed privacy decisions. |
| **2. Dynamic Evaluation** | Privacy Calculus Model (e.g., Acquisti et al., 2015) | A dominant cognitive process that actively reshapes the mental model. Users subconsciously perform a cost-benefit analysis, weighing the immediate utility of a service against abstract privacy risks. This often leads to acceptance of weaker privacy for perceived greater benefit. |
| | User Expectations (e.g., Sun & Yang, 2021) | What users believe should happen to their data. These are highly contextual and dynamic. A core challenge is the misalignment between these expectations and the system's actual technical privacy guarantees, directly influencing the privacy calculus. |
| **3. User-System Engagement Manifestations** | User Trust (e.g., Mehrabi et al., 2021) | A multifaceted outcome shaped by the alignment of the above factors. Trust is built not just by privacy, but by technical reliability, transparent communication, user control, and ethical alignment. It is the bridge between user perception and their willingness to use the system. |
| | User Acceptance (e.g., Schomakers et al., 2022) | The final decision and outcome of the user's internal privacy calculus. If the perceived privacy risks (shaped by mental models, expectations, and trust) outweigh the benefits, user acceptance plummets. It is directly mediated by privacy concerns. |
| | User Experience (e.g., Distler et al., 2020) | The qualitative feel of interacting with the AI system, which is fundamentally defined by privacy perceptions. A privacy-invasive system creates a negative experience through anxiety and cognitive burden, while seamless privacy protection enhances it. |
| **4. Moderating Mechanisms** | Personalized Privacy Attitudes (e.g., Sun & Yang, 2021) | The recognition that privacy preferences are individual and context-dependent. The goal is to move beyond one-size-fits-all models by creating systems that can adapt to individual user profiles, though this poses significant scalability and efficiency challenges. |
| | Human Error (e.g., Acquisti et al., | A vulnerability stemming from cognitive biases and behavioral pitfalls. A human-centered design must be error-tolerant to mitigate these inevitable |

| Category | Human Factors | Explanation & Relationship to Other Factors |
|---|---|---|
| | 2015) | mistakes, which can lead to accidental oversharing of data. |

5.1 Cognitive Foundations

User mental models, cultural differences, and awareness collectively form the cognitive foundation for understanding privacy in AI. User mental models form the core of this foundation, representing the user's internal understanding of how an AI system uses their data. However, these models are often incomplete or inaccurate, leading users to agree to terms without fully comprehending how their data will be collected, processed, and used, and remain confused about consent (Rudolph et al., 2018). This challenge is compounded by the fact that there is seldom a one-size-fits-all solution for privacy design in AI systems, necessitating the collection of data from each individual to develop their personalized models.

Cultural differences significantly shape these mental models. Studies on social media behavior indicate that users from individualistic cultures (e.g., the U.S.) tend to express greater privacy concerns, adopt more protective measures, and expect higher levels of control over their personal data compared to those from collectivistic cultures (Krasnova & Veltri, 2010; Zhong et al., 2024). Beyond individualism and collectivism, other dimensions create variation. For instance, South Korea and Germany exhibit particularly strong privacy awareness on social media (Krasnova & Veltri, 2010). Cultural context further influences specific privacy priorities, as Indonesian users may associate data sensitivity with religious practices (e.g., prayer), while European users rarely mention such concerns (Sun & Yang, 2021). Demographic factors such as age also introduce differences in approaches to data sharing.

User awareness directly counters the formation of poor mental models and serves as a critical safeguard against common human errors. Enhancing the level of user understanding regarding privacy risks and AI functionalities through interactive tutorials, clear in-system feedback, and simulations is crucial for enabling informed privacy decisions (Wongso et al., 2024). Promoting effective human-AI collaboration through "human-in-the-loop" mechanisms allows users to monitor and, when necessary, correct or override AI decisions related to data handling, fostering a sense of agency and reinforcing overall system trustworthiness (Shneiderman, 2020).

5.2 Dynamic Evaluation

Users dynamically evaluate privacy trade-offs through cognitive processes that are highly contextual and shaped by their expectations.

The privacy calculus model describes a dominant cognitive process that actively reshapes the mental model, where users subconsciously perform a cost-benefit analysis, weighing the immediate utility of a service against abstract privacy risks (Acquisti et al., 2015). This often leads to acceptance of weaker privacy for perceived greater benefit and makes users less likely to demand stronger protections, as they internalize the calculus that immediate utility nearly always justifies the uncertain cost of their data.

User expectations, or what users believe should happen to their data, are highly contextual and dynamic, varying across age, gender, race, education, income, religious beliefs, cultural background, and health status (Sun & Yang, 2021). A core challenge is the misalignment between these expectations and the system's actual technical privacy guarantees, which directly influences the privacy calculus. This misalignment is exacerbated because users' privacy attitudes are dynamic across different scenarios and domains. For example, privacy concerns negatively impact perceptions of AI as a substitute for doctors, especially for sensitive health issues (Liu et al., 2025), while social media users may exhibit entirely different privacy expectations. Measuring these preferences is complicated by the difficulty of accounting for all contextual factors, and longitudinal studies tracking long-term privacy concerns and shifts in awareness remain rare due to cost and time constraints (Herington, 2020; Sun et al., 2024). Furthermore, a significant obstacle is that most users lack a clear understanding of the technology or hold misconceptions about how their data will be used (Kyi et al., 2024), which widens the gap between expectation and reality.

5.3 User-System Engagement Manifestations

The cognitive foundations and dynamic evaluations ultimately manifest in how users engage with AI systems, reflected in their trust, acceptance, and overall experience.

User trust is a multifaceted outcome shaped by the alignment of the above factors. It is built not just by privacy, but by technical reliability, transparent communication, user control, and ethical alignment (Mehrabi et al., 2021). Technical performance and reliability form the foundation of trust, with consistent accuracy and explainable decision-making processes directly influencing users' confidence in AI systems (Liao et al., 2020). Transparency in communication plays a pivotal role by clarifying data usage practices, openly acknowledging system limitations, and providing real-time feedback about AI processes to end users (Binns et al., 2018). User control and agency significantly impact trust, where features such as adjustable privacy settings, revocable consent mechanisms, and interruptible AI operations empower users (Amershi et al., 2019). Social and contextual factors also manifest in trust formation. Cultural norms create significant variability, with individualistic societies emphasizing personal control and transparency (Krasnova & Veltri, 2010). Prior user experiences with technology establish critical baseline expectations; negative encounters with data breaches or algorithmic bias reduce trust, while positive experiences with well-designed interfaces increase willingness to engage with new AI applications (Razaque et al., 2025). The domain sensitivity of data adds another layer of complexity, with healthcare applications demanding particularly high thresholds for accuracy and confidentiality (Liu et al., 2025). Theoretical frameworks like Uncertainty Reduction Theory (URT), which outlines the axioms, strategies, and stages through which people reduce uncertainty in initial interactions, offer a valuable lens for understanding how users navigate the unknowns of AI systems and build trust (Kramer, 1999).

User acceptance is the final decision and outcome of the user's internal privacy calculus. If the perceived privacy risks (shaped by mental models, expectations, and trust) outweigh the benefits, user acceptance plummets, and it is directly mediated by privacy concerns (Schomakers et al., 2022). Drawing from technology acceptance models, perceived usefulness and perceived ease of use are significantly influenced by privacy perceptions. A privacy-invasive system undermines its usefulness through anxieties over data misuse and compromises its ease of use with the cognitive burden of managing privacy settings or understanding complex data practices.

User experience is the qualitative feel of interacting with the AI system, which is fundamentally defined by privacy perceptions (Distler et al., 2020). A privacy-invasive system creates a negative experience through anxiety and cognitive burden, while seamless privacy protection enhances it. Well-designed privacy features become invisible enablers of a positive experience, seamlessly integrating protection into user workflows while building trust through contextual transparency.

## 5.4 Moderating Mechanisms

Individual differences and cognitive limitations act as critical moderating mechanisms that influence the entire privacy perception landscape.

Personalized privacy attitudes recognize that privacy preferences are individual and context-dependent, with the goal of moving beyond one-size-fits-all models by creating systems that adapt to individual user profiles (Sun & Yang, 2021). However, scaling these personalized solutions presents significant challenges. As illustrated in Figure 4, these core challenges include scenario-specific limitations, evolving user awareness over time, divergent privacy expectations across different domains which make standardization difficult, and the technical complexity of managing numerous models without compromising system efficiency and utility (Wei et al., 2023). Furthermore, ensuring scalability across regions is challenging due to varying privacy regulations. The overarching goal for scaling privacy preference, therefore, must focus on achieving adaptability, usability (Teltzrow & Kobsa, 2004), interoperability, and regulatory alignment (Owen & Mattews, 2024). Promising research trajectories to reach these goals involve developing AI-driven, self-learning privacy models capable of adapting to dynamic behavioral patterns, conducting more extensive longitudinal studies examining the evolution of user privacy awareness, and designing flexible, portable privacy management frameworks capable of accommodating diverse regulatory environments.

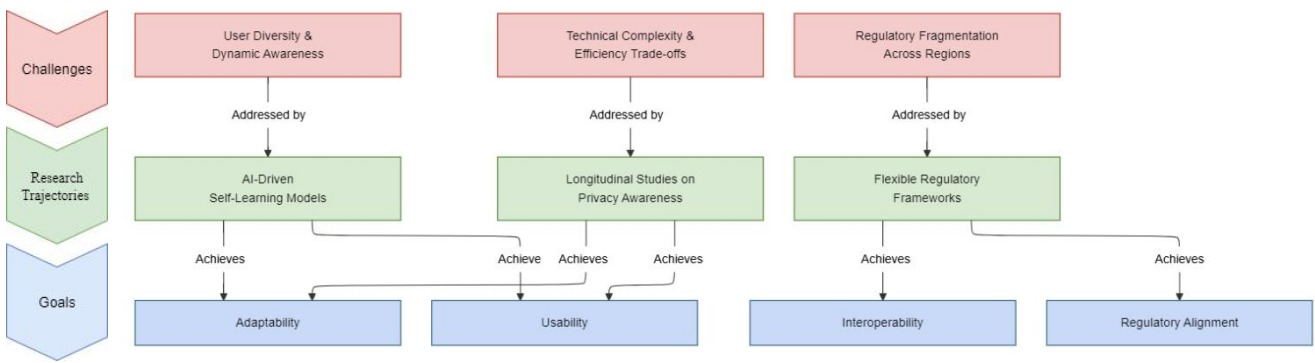

Fig 4. Challenges, Research Trajectories, Goals of Personalized Privacy Preferences

Human error remains a critical vulnerability stemming from cognitive biases and behavioral pitfalls (Acquisti et al., 2015). Users may accidentally overshare data due to misclicks, misunderstand risk due to optimism bias, or habitually accept consent prompts without reading them. A human-centered design must therefore be error-tolerant to mitigate these inevitable mistakes. This includes designing for error tolerance by providing clear undo functions (e.g., the ability to retract consent easily), implementing confirmation steps for critical actions, and structuring choice architectures to "nudge" users toward safer privacy behaviors without compromising their autonomy (Soh, 2019).

## 6. Privacy Regulations

The previous section established that trust is eroded when a disconnect exists between technical systems and user expectations. This gap cannot be bridged by technical design and ethical intent alone. Robust guardrails are required. This section examines the regulatory and ethical frameworks that translate abstract principles of AI privacy into concrete action. It investigates how evolving global regulations, from General Data Protection Regulations (GDPR) to emerging AI acts, attempt to codify protections for mental privacy, personal autonomy, and trust. This section then focuses on a pivotal right: the "right to explanation" and its profound implications for building human-centered and understandable AI systems. Finally, it concludes by addressing the challenge of aligning ethical guidelines with regulatory compliance and user values.

6.1 Overview of Global Data Protection Laws

The protection of personal data has become a critical concern worldwide, leading to the development of comprehensive data protection laws across different jurisdictions. Two of the most significant regulations are the European Union's General Data Protection Regulation (GDPR) (European Union, 2016) and the United States' California Consumer Privacy Act (CCPA) (California Legislature, 2018).

In the European Union, the GDPR implemented in 2018, marked a transformative shift in data protection standards, superseding the 1995 Data Protection Directive (Directive 95/46/EC) (European Parliament and Council, 1995). This landmark legislation establishes robust privacy rights for individuals, including the right to be informed about data collection practices, to access personal data, to rectify inaccurate information, and to erasure, which is commonly known as the "right to be forgotten." (Bakare et al., 2024)

The California Consumer Privacy Act (CCPA) provides similar protections for residents of California. The CCPA grants consumers several fundamental rights: the right to know what personal information is being collected, to opt-out of the sale of their data, and to request deletion of their personal information. This regulatory framework was further expanded in 2023 with the introduction of the California Privacy Rights Act (CPRA) (Determann & Tam, 2020). The CPRA incorporates key principles from the GDPR, including data minimization, purpose limitation, and storage limitation. Notably, the CPRA also enhances consent requirements, bringing them closer to the strict standards employed in the European Union.

Apart from that, China's Personal Information Protection Law (PIPL) (National People's Congress of the People's Republic of China, 2021) represents the China's first comprehensive data privacy legislation. It partly modeled GDPR, such as consent and data subject rights, but establishes strict rules for processing personal data within China (data localization rules).

Notably, most of the European countries adapt EU GDPR and supplement rules in specific areas, such as Federal Data Protection Act (BDSG; Federal Ministry of Justice and Consumer Protection, 2023) in Germany, Data Protection Act (Loi Informatique et Libertés; French National Assembly, 2018) in France, and Personal Data Protection Code (Italian Parliament, 2018) in Italy.

Though there are different data protection laws, they have many common aspects, especially they regulate the core privacy rights, including the right to be informed about data collection practices, to access personal data, to rectify inaccurate information (correct inaccurate or incomplete personal data), to data portability, and to erasure (European Union, 2016). In addition, CCPA and CPRA has regulated the right to non-discrimination (Baik, 2020), which applies to all businesses under CCPA's jurisdiction.

The laws mentioned above and their key features are listed in the Table 4, the regulations in the table are selected based on influence and geographic representation.

**Table 4 Overview of Global Data Protection Laws**

| Regulation | Region | Key Features |
| --- | --- | --- |
| General Data Protection Regulation (GDPR) (European Union, 2016) | EU | Strict consent, broad rights |
| California Consumer Privacy Act (CCPA) (California Legislature, 2018) & CPRA (California Legislature, 2020) | California, USA | Opt-out of data sales |
| Personal Information Protection Law (PIPL) (National People's Congress of the People's Republic of China, 2021) | China | Consent, data localization |
| ePrivacy Directive (Cookie Law) (European Parliament and Council, 2002) | EU | Electronic communications privacy, consent for non-essential cookies |
| Federal Data Protection Act (BDSG) (Federal Ministry of Justice and Consumer Protection, 2023) | Germany | Strict rules on employee data and video surveillance |
| Data Protection Act (French National Assembly, 2018) | France | Special rules on biometric data and credit scoring |
| Personal Data Protection Code (Italian Parliament, 2018) | Italy | Additional rules on electronic marketing and data breach notifications |

6.2 The "Right to Explanation" and Its Implications of Human-Centered AI

The "right to explain" is a legal and ethical principle ensuring that individuals affected by AI systems are entitled to clear, meaningful, and actionable explanations for automated decisions. This right upholds three fundamental principles. First, users must be informed when and how an AI system influences decisions about them (e.g., loan approvals, hiring, or medical diagnoses), which ensures transparency. Second, the AI's decision-making logic should be understandable to humans, avoiding "black-box" obscurity, which ensures interpretability. Third, organizations deploying AI must justify automated outcomes and provide redress mechanisms for erroneous or biased decisions, which ensures accountability (Yeung, 2020; Ribeiro et al., 2016). Since human-centered AI prioritizes human control, trust, accountability, and fairness, ensuring transparency and interpretability in AI systems is essential (Xu, 2024). By making AI decisions understandable to users, trust is enhanced in the system. Additionally, clear explanations help detect and mitigate biases, reducing the risk of discrimination in automated decision-making.

6.3 Organizational Responsibility, Data Stewardship, Data Governance Mechanisms

Organizational responsibility, data stewardship, and data governance are interconnected yet distinct concepts in ethical AI management. Organizational responsibility establishes the overarching ethical and legal obligations for AI systems, focusing on principles like fairness and accountability. Data stewardship operationalizes these principles through hands-on data management practices such as anonymization and consent protocols throughout the data lifecycle (Jacobsen et al., 2020). Data governance provides the structural foundation through policies, access controls, and compliance mechanisms that enable effective stewardship. While governance creates the framework, stewardship implements daily practices, and organizational responsibility guides the ethical vision, ensuring ethical principles are embedded in deployment.

All three share common goals of ensuring regulatory compliance, maintaining data integrity, and building trustworthy AI systems (Gudepu et al., 2024). Their interdependence is evident in practice. For instance, governance policies empower stewards to maintain quality data, which in turn helps organizations fulfill their responsibility to deploy unbiased AI. Together, they form a comprehensive approach to responsible data and AI management, with governance as the infrastructure, stewardship as the execution, and organizational responsibility as the guiding ethos.

6.4 Aligning Ethical Guidelines with Regulatory Compliance and User Values

Aligning regulatory compliance with user values presents significant difficulties due to tensions between legal requirements, ethical principles, and societal expectations. Regulatory frameworks often prioritize risk mitigation and legal enforceability, while user values emphasize transparency, fairness, and usability. For example, GDPR imposes strict consent requirements that can conflict with user desires for seamless digital experiences (Wachter et al., 2017). Additionally, cultural variations in privacy expectations create complications for global systems. What satisfies European regulators may not align with user expectations in other regions (Zuboff, 2019). The rapid evolution of AI technologies further exacerbates this gap, as regulations struggle to keep pace with innovation (Cath, 2018).

To achieve alignment between regulatory compliance and user values, it is effective to enable human-AI collaboration and actively incorporate user perspectives when designing AI systems. For organizations, moving beyond mere legal adherence to foster trust by embedding ethical principles, such as fairness, accountability, and transparency, into their culture and operations is a practical solution.

7. **Privacy Governance**

Effective privacy governance in the context of AI demands a holistic and proactive approach that systematically addresses potential privacy risks across the entire AI lifecycle, spanning from the initial stages of data collection and processing to subsequent model deployment, feedback loop integration, and long-term data reuse. Given the dynamic and multi-faceted nature of AI systems, privacy governance initiatives cannot be fragmented or limited to isolated phases; instead, they must be embedded as a core component of each stage in the lifecycle to effectively mitigate the specific privacy risks identified in Section 2, such as, re-identification, algorithmic bias, and non-compliance with regulatory requirements. By integrating governance measures throughout the lifecycle, organizations can ensure that privacy is not an afterthought but a foundational principle guiding AI development and operation.

7.1 Data Governance Across AI Lifecycle Stages

To establish robust privacy governance, it is critical to implement targeted measures at each distinct stage of the AI lifecycle, starting with the initial data collection phase. At this juncture, governance mechanisms should prioritize and enforce three fundamental privacy principles: data minimization, purpose limitation, and contextual integrity. To operationalize these principles, organizations are required to deploy systematic consent management frameworks that not only capture user permissions at the point of data collection but also maintain transparent and auditable tracking of these permissions throughout the entire data lifecycle, from storage and processing to potential sharing or reuse (Zaeem & Barber, 2020). Complementing these organizational frameworks, technical controls play a pivotal role in validating that only essential data points are collected, thereby reducing the volume of sensitive information at risk. Concurrently, organizational policies must be formalized to ensure full transparency regarding data usage purposes, including clear communication to users about how their data will be processed, who will have access to it, and the duration of its retention (Mantelero & Esposito, 2021).

Moving to the data processing phase, privacy governance mechanisms must mandate the adoption of privacy-preserving techniques to safeguard data while enabling legitimate AI operations. Key techniques in this regard include differential privacy, k-anonymity, and secure multi-party computation (Abadi et al., 2016). Additionally, organizations need to establish Data Protection Impact Assessments (DPIAs) that are specifically tailored to the unique characteristics of AI systems. Unlike generic DPIAs, these assessments should proactively evaluate risks associated with AI-specific processing activities, such as the potential for re-identification of anonymized data, unintended aggregation of disparate datasets, and privacy harms stemming from algorithmic decisions, before processing commences. This proactive risk evaluation is not only a best practice but also a regulatory requirement under frameworks such as the General Data Protection Regulation (GDPR, 2016), which mandates DPIAs for high-risk processing activities (Kamarinou et al., 2016).

## 7.2 Adaptive Governance

A critical limitation of static privacy governance frameworks is their inability to keep pace with the rapid evolution of AI technologies, emerging privacy threats, and shifting regulatory landscapes. Thus, privacy governance cannot be treated as a one-time implementation. Instead, it requires continuous monitoring, evaluation, and refinement throughout the entire AI lifecycle. To achieve this, organizations should deploy real-time privacy impact monitoring systems that track data flows, usage patterns, and algorithmic outputs in real time, enabling the timely detection of anomalies or non-compliant activities (Shastri et al., 2019). Regular privacy audits and assessments are also essential, particularly during critical junctures such as model updates, retraining, or scaling of AI systems, moments when new privacy risks may emerge due to changes in data inputs, algorithmic parameters, or operational contexts (Veale et al., 2018). Furthermore, governance mechanisms must address the unique risks posed by AI feedback loops, where data generated by the AI system (e.g., user interactions, model outputs) is reused to refine the system. This requires implementing robust data provenance tracking and consent refresh protocols(Binns et al., 2018).

In essence, effective privacy governance must be adaptive, characterized by a feedback-driven approach that incorporates lessons learned from privacy incidents, emerging research on AI vulnerabilities, and evolving regulatory expectations. This adaptability is particularly crucial for addressing threats that continue to emerge in AI deployment, such as membership inference attacks and model inversion attacks (Hayes et al., 2019). By fostering a culture of continuous improvement and embedding adaptability into governance mechanisms, organizations can ensure that their privacy measures remain effective in mitigating both existing and emerging risks, thereby upholding user trust and regulatory compliance in the long term.

## 8. HCAI Design Guidelines for Human-Centered Privacy

Design guidelines are voluntary best practices, which are upheld by a desire for quality, peer influence, and market forces rather than by law. Expert bodies and industries often benefit more from these guidelines, as they provide actionable advice on how to achieve effective design and a superior user experience. Section 8 summarizes design guidelines for AI privacy from the Human-Centered perspective. It is crucial to distinguish these from design principles. Design principles are broad, foundational statements that guide the overall strategy and philosophy of a design. They answer the question of "what" to achieve, such as "ensure user control" or "minimize data collection." In contrast, design guidelines are specific, actionable recommendations derived from these principles. They answer the question of "how" to implement the principles in practice. For instance, the principle "ensure user control" might lead to a guideline stating, "provide a clear and accessible dashboard where users can view, edit, and delete their stored data." Therefore, while principles provide the strategic direction, guidelines offer the concrete, tactical steps for execution, making them more directly applicable for designers and developers (Schoormann et al., 2024). Table 5 presents design principles and design guidelines, with illustrative examples provided for each guideline. The following sections discuss the design guidelines derived from these design principles.

**Table 5 HCAI Design Guidelines, Principles, Examples**

| HCAI Design Principles | Design Guidelines | Example |
|---|---|---|
| **Value-Sensitive Design** | Center privacy design around moral values such as autonomy, dignity, and respect for individuals' rights. | Apple's "App Tracking Transparency" feature highlights ethical framing by defaulting to user opt-out (Friedman & Hendry, 2019) |
| **Explainable Privacy** | Provide intelligible, contextual information about data practices at key moments. | Google's "My Activity" and YouTube's "Why this ad?" features surface personalized data flows (Kizilcec, 2016) |
| **Granular Control** | Go beyond binary permissions; provide layered controls for different data types or use cases. | Instagram allows users to control visibility by post and restrict audience by story (Luger & Rodden, 2013) |
| **Participatory Design** | Involve users early in the privacy design process to reflect lived realities and foster trust. | Mozilla engaged users in shaping its Enhanced Tracking Protection feature (Iachello & Hong, 2007) |
| **Inclusive and Culturally Sensitive Design** | Respect diverse cultural expectations, languages, and mental models in privacy UX. | WhatsApp defaults to E2E encryption, requiring no opt-in—effective in multilingual, global contexts (Awad & Krishnan, 2006) |
| **Human-Centered Evaluation of Privacy Technologies** | Evaluate AI privacy using human factors like understandability, usability, and comfort (not just mathematical rigor). | Apple's public white papers on privacy enable user trust and transparency beyond the math of differential privacy (Shokri et al., 2017) |
| **Contextual Privacy Defaults and Just-In-Time Consent** | Design interfaces that ask for data access when most relevant, using conservative defaults. | Fitbit activates location access only when GPS-based activity is initiated (Shneiderman, 2022) |
| **Progressive Disclosure** | Avoid overwhelming users. Present privacy information gradually as needed, using progressive layering. | Facebook provides summaries of data settings with "learn more" links for advanced users (Eiband et al., 2019) |
| **Mental Model Alignment** | Design interactions that fit users' intuitive beliefs about how data flows and AI systems work. | Smart speaker privacy toggles shaped like physical mute buttons build intuitive assurance (Kang et al., 2015) |
| **Privacy Nudges** | Use subtle design cues to encourage privacy-protective behaviors without restricting freedom. | Browser warnings about unsafe sites or third-party trackers nudge users without denying access (Acquisti et al., 2015) |

## 8.1 Value-Sensitive Design

To mitigate potential harms in an era of advanced technology and ensure that information technology operates ethically, researchers have proposed value-sensitive design (Sadek et al., 2024). This approach integrates moral values into the design, research, and development of ICT, ensuring that ethical considerations shape technological innovation. Value-sensitive design emphasizes that privacy should not be seen merely as a technical problem but as a moral and human value deeply tied to individual dignity and the ability to act autonomously. It centers privacy design around moral values such as autonomy, dignity, and respect for individuals' rights.

## 8.2 Explainable Privacy

Explainable privacy refers to the practice of clearly explaining how each privacy-preserving mechanism contributes to protecting user privacy within a specific application design context. This guideline is to ensure that privacy policies, data-handling practices, and system behaviors are not hidden, confusing, or overly technical. Instead, they should be communicated in a way that is clear and transparent, enabling users to make informed decisions. Providing contextual information helps both users and designers understand what data is being collected, how it flows through the system, and who has access to it. By making these practices transparent and interpretable, privacy becomes not only a technical safeguard but also something that is comprehensible and understandable to all stakeholders (Alkhariji et al., 2021).

## 8.3 Granular Control

Granular control in privacy management systems enables hierarchical user access to the data, allowing different users to access varying levels of data granularity within the same dataset, providing users with precision tools to manage their digital footprint (Khan et al., 2021). This guideline respects individual users' conditions and preferences, emphasizing that privacy controls should go beyond simple all-or-nothing choices (e.g., "share everything" vs. "share nothing"). Rather, privacy controls should provide fine-grained, flexible, and context-sensitive options. Users ought to decide what data is shared, with whom, when, and under what conditions—in ways that reflect their personal needs and circumstances.

## 8.4 Participatory Design

Participatory design is an approach to computer system development that actively involves end-users as key contributors throughout the design process, reflecting lived realities and foster trust (Wacnik et al., 2025). This design approach is unlike user-centered design, where designers lead the process on behalf of users based on research findings. It actively transfers agency to users, involving them in the co-creation of prototypes, design concepts, and final solutions (Simonsen & Robertson, 2013). In the process of participatory design, instead of designers and engineers making assumptions about what users need for their privacy, users are brought in as partners to co-design the features that will protect them. This design guideline enables users to design on their own, allowing personalized design instead of one-fits-all solutions.

## 8.5 Inclusive and Culturally Sensitive Design

This design guideline advocates for respecting users with diverse functional needs and cultural backgrounds. It integrates two complementary approaches: inclusive design, which ensures usability and access for people across the full spectrum of abilities and disabilities, and culturally sensitive design, a framework that prioritizes respect for diverse cultural expectations, linguistic nuances, and localized mental models.

As Lee & Sayed (2008) note, designing for a global audience is a critical challenge in contemporary practice. This is particularly salient in privacy UX, where this approach addresses variations in perceptions of data ownership, rituals of consent, and trust-building practices. By ensuring digital products align with users' cultural contexts, culturally sensitive design enhances adoption and legitimacy across different regions (Lee & Sayed, 2008).

8.6 Human-Centered Evaluation of Privacy Technologies

Human-centered evaluation marks a paradigm shift in assessing the effectiveness of PETs. While traditional analysis has focused on technical metrics, such as cryptographic strength, algorithmic efficiency, and theoretical privacy guarantees—this approach crucially expands the criteria to incorporate human factors. Traditionally, the user has often been characterized as the weakest link in the security chain, with many recent data breaches attributed to human error, whether deliberate or accidental. Consequently, integrating human factors into technology design is critical (Nepal et al., 2022). A PET's true efficacy, therefore, depends not only on its technical robustness but also on its usability. Evaluation must examine whether users can readily understand the tool's functionality, configure it correctly, and trust its protections. Ultimately, success is measured by the ability to integrate the technology into daily routines without committing errors that compromise privacy (Groen et al., 2023).

8.7 Contextual Privacy Defaults and Just-In-Time Consent

Although comprehensive privacy policies can address most privacy concerns and safeguard users' information, a practical challenge exists: people are now confronted with hundreds of privacy-related decisions, leading to sporadic and limited attention to lengthy privacy policies. To address this, contextual privacy defaults and "just-in-time consent" (decision-making based on "just-in-time notice") have been introduced. These approaches provide users with prompts to grant or deny data access at the most relevant moment, while applying conservative defaults in other cases (Soeder, 2019).

The idea behind "just-in-time notice" is to support informed consent by asking for it when it matters most. At the exact moment of relevance, a timely warning can make users recognize the importance of the data being requested, prompting them to reconsider their choices (Soeder, 2019). This not only saves time but also helps users make more efficient and thoughtful decisions.

8.8 Progressive Disclosure

Progressive disclosure involves hiding advanced interface controls to reduce initial user errors and facilitate learning. In essence, complex information and explanations are presented only upon request, serving users on an "as-needed" basis. This design guideline aligns with insights from explanation research, which suggests that explanations are typically occasioned in human-human interactions, provided when the situation demands (Chen et al., 2024) and can be applied to enhance transparency in intelligent systems (Springer & Whittaker, 2018).

8.9 Mental Model Alignment

Mental model alignment refers to designing interfaces that reflect users' existing expectations, knowledge, and cognitive frameworks. When a system behaves in line with these mental models, interactions feel more intuitive, reducing errors and cognitive effort. This approach helps users focus on their goals rather than deciphering how the tool works, significantly enhancing learnability and usability (Young, 2008).

8.10 Privacy Nudges

A privacy nudge is a subtle design intervention that guides users toward more informed privacy decisions without restricting their choices or relying on complex notifications. Grounded in behavioral economics, nudges offer an alternative regulatory mechanism to traditional informed consent models (Soh, 2019; Rees et al., 2022). Privacy nudges are not meant to replace the existing notice-and-choice framework but to enhance it by providing flexible, effective, and user-friendly privacy protections. Unlike conventional privacy notices, nudges leverage direct product interactions to deliver timely, intuitive, and contextually embedded guidance (Calo, 2012; Monteleone et al., 2015).

9. Case Studies and Applications

Section 9 explores the intersection of AI and personal privacy across key sectors of society. As AI systems become more integrated into daily services, they bring benefits in efficiency, personalization, and security, but also pose privacy challenges that require careful examination of how sensitive data is collected, utilized, and safeguarded. The section covers healthcare, education, finance, and smart cities, and concludes with participatory and user-centered approaches to AI audits and evaluations. The aim is to empower users, integrate diverse perspectives, and ensure privacy protections are embedded by design. Table 6 provides the summary of AI privacy applications and challenges across sectors.

Table 6 Summary of AI Privacy Applications and Challenges Across Sectors

| Domain | Key AI Applications | Core Privacy Risks & Challenges | Case Study |
|---|---|---|---|
| Healthcare | Disease prediction, treatment optimization, administrative automation | Unauthorized access to sensitive patient data, lack of proper consent, insufficient data impact assessments | Google DeepMind and NHS (2017): Accessed 1.6M patient records without proper consent for the Streams app. |
| Education | Tailoring coursework, plagiarism detection, remote exam proctoring | Excessive data collection (e.g., biometrics), intrusive surveillance, risk of data misuse by third parties | Proctorio (2020): Backlash over intrusive surveillance during exams.<br><br>NYC Public Schools (2023): Banned ChatGPT over data misuse fears. |
| Finance | Customer segmentation, personalized services, spending analysis | Algorithmic bias leading to discrimination, lack of transparency, invasive data harvesting without explicit consent | Klarna (Ezennaya-Gomez et al., 2022) & Revolut (Polasik et al., 2022): Criticized for analyzing shopping behavior and spending habits to push services/ads, raising over-indebtedness and consent concerns. |
| Smart Cities | Facial recognition, predictive policing, IoT sensor networks | Mass surveillance, data misuse, algorithmic bias, violation of privacy rights and human autonomy | London Metropolitan Police (Dodd et al., 2022): Use of Live Facial Recognition (LFR) was ruled to violate human rights due to its privacy impact (~70% accuracy). |
| Design | Ensuring user engagement, ensuring transparency | AI systems can perpetuate bias if not tested with diverse users, leading to unfair or harmful outcomes | Google's PAIR Initiative (McAran, 2021): Used User-Centered Design (UCD) to test its AI chatbot, uncovering biases in handling sensitive topics like mental health. |

9.1 AI Privacy in Healthcare

Healthcare is undergoing a digital transformation with AI-driven innovations in electronic health records (EHRs), wearable devices, and telemedicine. While these technologies improve patient care, they also introduce significant privacy risks requiring robust safeguards to protect sensitive health data. Hospitals and nursing homes face cybersecurity threats and hacking incidents.

The 2017 collaboration between Google DeepMind and the UK's Royal Free NHS Foundation Trust focused on developing Streams, an AI-powered application to analyze patient data and detect early signs of acute kidney injury. However, privacy concerns arose when it was revealed that DeepMind had accessed 1.6 million patient records without proper consent. The UK Information Commissioner's Office determined that this data sharing violated the Data Protection Act, citing a lack of lawful basis for processing and the absence of a thorough data privacy impact assessment. In response, DeepMind implemented reforms including independent audits of data access, a patient data oversight panel, and more transparent consent procedures. The controversy contributed to the integration of DeepMind Health into Google Health in 2018, raising continued concerns about corporate access to sensitive medical data. This episode remains a landmark reference in discussions on ethical AI deployment and the importance of maintaining public trust in health data partnerships (BBC News, 2017). As public trust is the most critical component of any successful innovation, without prioritizing privacy-by-design and ethical governance, technological progress risks will be undermined by a failure of legitimacy.

9.2 AI Privacy in Education

AI is increasingly integrated into educational systems through personalized learning platforms, automated grading, and student monitoring tools. Platforms such as Carnegie Learning and Knewton (Demianenko, 2019). personalize coursework, while Turnitin (Meo & Talha, 2019) uses machine learning for plagiarism detection. These technologies hold promise for improving learning experiences but raise important privacy concerns related to the collection and analysis of sensitive student data. A notable 2020 case involved Proctorio, an AI-powered remote exam proctoring platform, which faced backlash for excessive data collection, including biometric data and surveillance practices (News, 2020). In 2023, NYC Public Schools banned ChatGPT due to concerns over data privacy (Ropek, 2023).These cases also demonstrate that public trust, once broken by privacy overreach, is far more difficult to restore than any technical systems.

8.3 AI Privacy in Finance

The finance sector relies on AI for credit scoring, fraud detection, and consumer profiling. While AI improves efficiency and accuracy, it raises data privacy concerns due to the sensitive nature of financial information. Regulatory frameworks such as GDPR (EU) and FCRA (US) impose strict requirements on data handling. Unauthorized access, algorithmic bias, and lack of transparency can lead to discrimination, identity theft, and loss of consumer trust. Financial institutions use AI to segment customers and personalize services, but concerns about invasive data harvesting persist. Case examples include Klarna (criticisms over debt risks and privacy concerns) and Revolut (criticisms over tracking spending habits for targeted ads). CPRA now mandates disclosure of AI-driven profiling and opt-out rights for users (Determann & Tam, 2020).

9.3 AI Privacy in Smart Cities and Public Surveillance

The deployment of AI-driven public surveillance technologies in smart cities, such as facial recognition, predictive policing, and IoT sensor networks across the cities, has sparked intense debate over the trade-offs between public safety and individual privacy and autonomy. While these systems are designed to improve urban security, traffic management, and emergency response, they also raise significant concerns regarding mass surveillance, data misuse, and algorithmic bias. Facial recognition for crime prevention offers a telling example. In London, the Metropolitan Police employs live facial recognition cameras to scan crowds for suspects, reporting an accuracy rate of around 70% (Dodd et al., 2022). However, in 2020 a court ruled that this practice violated privacy rights protected under the European Convention on Human Rights (Aykhan, 2023). This case highlights the urgent need for a well-regulated framework that ensures both the privacy of individuals and the effectiveness of public surveillance.

9.4 Participatory and User-Centered Approaches in AI System Audits and Evaluations

Research demonstrates that implementing user-centered design (UCD) in AI applications significantly enhances usability and user satisfaction. By prioritizing user needs, UCD improves system performance metrics, ensures compliance with human-centric design principles, and increases user engagement and adoption rates (Wongso et al., 2024). Compared to user-centered design which prioritizes users' feedback but is lead by experts, participatory design actively involves end-users as co-creators. A notable case from Google's People + AI Research (PAIR) applied UCD principles to evaluate AI chatbot fairness, revealing biases in handling sensitive topics, leading to redesigns and the development of a public transparency framework (McAran, 2021).

# 10. Challenges and Future Directions

This section discusses challenges and solutions for integrating privacy into AI systems, focusing on explainable privacy, human-in-the-loop privacy monitoring, scaling personalized privacy preferences, and addressing privacy concerns in generative AI and large language models. It emphasizes the need for multidisciplinary research to embed privacy in human-centered AI ecosystems and to pursue privacy-aware innovation.

## 10.1 Explainable Privacy

As AI systems increasingly handle sensitive user data, ensuring transparency and privacy has become critical. Traditional privacy mechanisms (e.g., differential privacy, encryption) rely on mathematical guarantees that are often incomprehensible to non-experts, creating a gap between technical safeguards and user trust (Erotokritous et al., 2024). Also, providing users with technical guarantees in detail is overwhelming, while simplified information will mislead users and cause wrong decisions (Asthana et al., 2024). Making technical privacy guarantees human-comprehensible is therefore urgent.

For future improvements in making privacy guarantees explainable, human-centered design should be adopted. Such an approach benefits users by prioritizing their needs and making abstract technical guarantees more intuitive through clear text and visual methods. From a design perspective, interactive tools can also help users explore different scenarios and understand the consequences of each decision. In addition, privacy guarantees could be summarized in standardized labels that remain consistent across platforms, making them more portable and applicable in diverse fields. Finally, personalized explanations can be provided to adapt to users' varying levels of knowledge and background (Dammu et al., 2021).

## 10.2 Human-In-The-Loop Privacy Monitoring and Assurance

As AI systems process vast amounts of personal data, fully automated privacy controls often fail to address contextual nuances, ethical dilemmas, and evolving threats. Human-in-the-loop (HITL) privacy integrates human oversight with AI-driven monitoring to detect unintended data misuse (e.g., biased inferences, re-identification risks) (Holstein et al., 2019), resolve edge cases where algorithmic privacy checks are insufficient (e.g., medical data sharing in emergencies) (Price & Cohen, 2019), and provide accountability for compliance with GDPR, CCPA, and other regulations (Wachter et al., 2021).

Taking healthcare as an example, even when patients' or elderly individuals' privacy preferences are collected in advance to guide an AI system's automated decisions, there will inevitably be fuzzy boundaries and exceptional cases. Privacy is not static, as patients' awareness and comfort levels evolve over time, especially when faced with sensitive or urgent medical situations. For instance, a patient who normally restricts data sharing may wish to allow broader access during an emergency to receive timely medical help. In such scenarios, relying solely on automated health monitoring could lead to false predictions or unnecessary alarms. Introducing human-in-the-loop monitoring allows clinicians or caregivers to intervene when unusual cases arise, ensuring that privacy preferences are respected while still prioritizing patient safety. This hybrid approach not only reduces false alarms but also builds trust, as patients know that critical decisions are not left entirely to machines.

10.3 Privacy in Generative AI and Large Language Models

The rapid advancement of generative AI (GenAI) and large language models (LLMs) has unlocked powerful capabilities in content creation, automation, and decision-making. Yet, these technologies also pose significant privacy challenges, as they often rely on vast amounts of training data that may include sensitive or personal information (Carlini et al., 2021; Weidinger et al., 2021). As GenAI and LLMs become increasingly pervasive, addressing these challenges is essential for ethical deployment, regulatory compliance (European Union, 2024), and maintaining user trust. Future work must carefully balance model performance with robust privacy preservation, while also promoting accountability and transparency in AI systems. Established techniques such as differential privacy, federated learning, and homomorphic encryption can be adapted, and potentially revolutionized, to meet the evolving demands of these advanced AI models.

10.4 Multidisciplinary Research for Human-Centered Privacy Framework

The development of human-centered AI systems that genuinely respect privacy necessitates a comprehensive, multidisciplinary approach. Privacy in AI represents a complex interplay between technological capabilities and human values, requiring collaboration across multiple fields (Zuboff, 2019). Computer science provides the foundational algorithms and security architectures, such as differential privacy (Dwork & Roth, 2014). Law and policy establish the regulatory frameworks governing data use (European Union, 2016; California Legislature, 2018). Ethics and philosophy contribute critical perspectives on fairness and autonomy.

Critically, human factors and psychology are indispensable for bridging the gap between technical systems and user trust, with a particular emphasis on mitigating the pervasive risks of human error and behavioral biases. They offer insights into human decision-making processes (Acquisti et al., 2015), the formation of mental models (Kang et al., 2015), and the determinants of user acceptance, which are essential for designing comprehensible and usable privacy controls. This field addresses vulnerabilities such as accidental misclicks, habitual consent without comprehension, and optimism bias in risk assessment (Acquisti et al., 2015). Furthermore, human factors underscore that technology alone is insufficient. It must be coupled with enhanced user awareness and "human-on-the-loop" oversight. Training, interactive feedback, and literacy initiatives (Wongso et al., 2024) empower users to correct AI behavior and make informed decisions, transforming them from passive subjects into active collaborators in privacy protection. This human-centric perspective is complemented by the social sciences, which illuminate cultural variations and societal impacts (Acquisti et al., 2020; Krasnova & Veltri, 2010; Sun & Yang, 2021). Only through such multidisciplinary integration can AI ecosystems be developed that are both technically robust and socially responsible (Crawford, 2021).

Figure 5 provides a comprehensive view over multidisciplinary approach required for developing privacy-preserving human-centered AI ecosystems. By fostering deep collaboration between technologists, social scientists, legal experts, and end-users, researchers can develop AI systems that not only protect data but also respect human dignity, cultural values, and individual autonomy. Future research must prioritize this holistic integration to create trustworthy AI ecosystems capable of navigating our complex digital society.

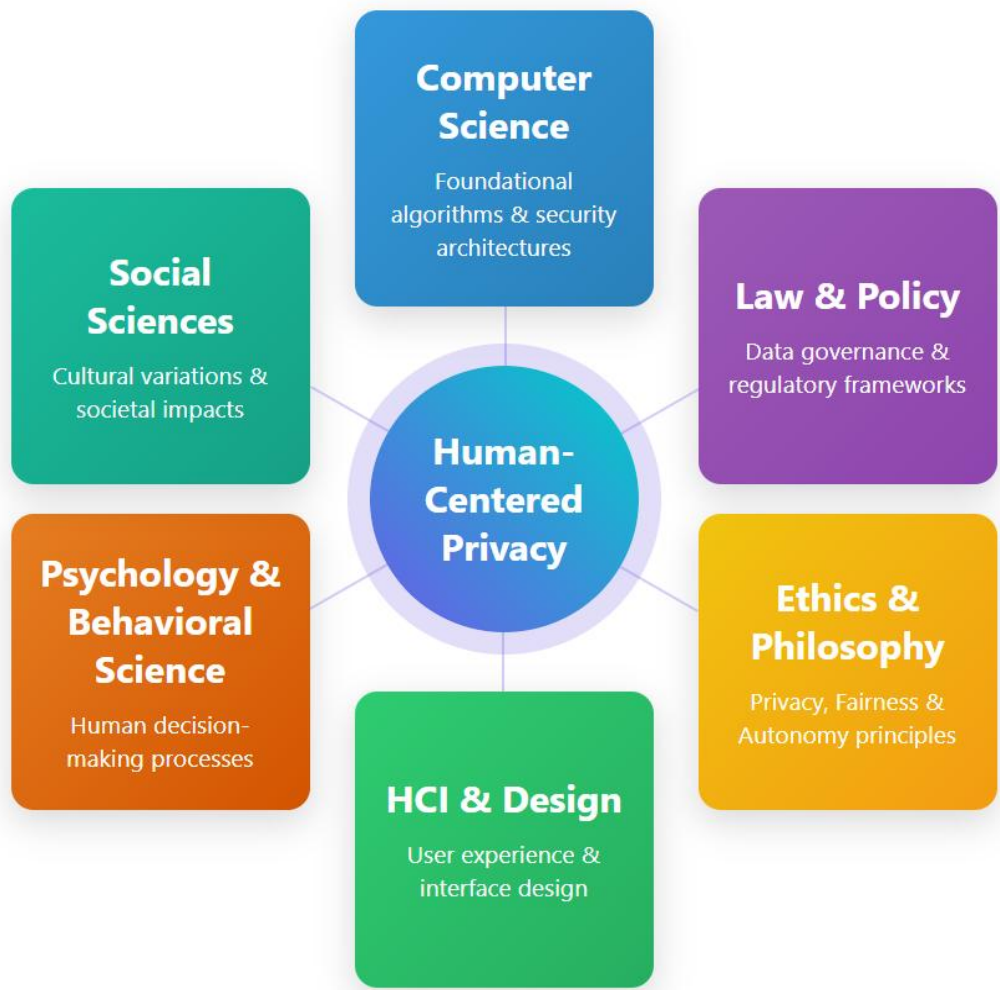

Fig 5. Multidisciplinary Collaboration for Human-Centered Privacy

10.5 AI Innovation/Development and AI Privacy

The relationship between AI technology and AI privacy is inherently a balancing act. On one hand, overemphasizing privacy protections may constrain data availability, limit model performance, and slow down innovation. On the other hand, neglecting privacy considerations creates long-term risks, such as data breaches, misuse of sensitive information, loss of user trust, and regulatory backlash, which can undermine the sustainability of AI technologies. This challenge is not a zero-sum trade-off but an opportunity to pursue privacy-aware innovation. Approaches such as DP, FL, and secure computation show that it is possible to design systems that preserve individual privacy while still enabling AI progress. The key lies in developing scalable, efficient, and explainable privacy-preserving methods that integrate seamlessly into AI pipelines.

**11. Conclusion**

The concluding section synthesizes the critical arguments presented throughout the article, emphasizing that privacy is fundamental in human-AI systems and requires multidisciplinary collaboration. It moves beyond mere summary to present a forward-looking vision, charting a path for integrating privacy as a core value into the very fabric of human-AI collaboration in future design.

## 11.1 Recap of Importance of Human-Centered Privacy

The pervasive influence of AI is no longer a futuristic concept but a daily reality, evidenced by the widespread adoption of tools like ChatGPT and DeepSeek. As these technologies become increasingly embedded in our lives, privacy protection emerges as crucial for safeguarding individual autonomy, dignity, and trust in AI systems. Privacy in AI transcends mere legal or technical compliance. It is a foundational ethical principle for responsible, human-centered privacy design. By protecting personal data, researchers not only mitigate risks of misuse, discrimination, and surveillance but also cultivate greater user confidence and engagement with AI technologies. This human-centered approach aligns technological advancement with fundamental human values, making privacy protection indispensable to this mission.

Practical implementation occurs at multiple levels. First, empowering users through "data minimization" habits, encouraging avoidance of sensitive personal information in public AI platforms. Second, integrating advanced privacy-preserving techniques like federated learning, differential privacy, homomorphic encryption, and secure multi-party computation. These approaches ensure AI systems remain accountable and fair while harnessing AI's benefits without compromising the fundamental right to privacy.

## 11.2 The Role of Design, Policy, and Collaboration in Advancing Privacy-Aware AI

Advancing privacy-aware AI requires a holistic and synergistic framework that integrates design, policy, and cross-disciplinary collaboration. These elements are deeply interconnected and mutually reinforcing. From a design perspective, privacy must be embedded into the architecture of AI systems through technical measures such as federated learning and differential privacy, as well as design principles like data minimization and privacy by default. At the policy level, robust legal and normative foundations are essential, including clear data governance standards, algorithmic accountability mechanisms, and adaptive compliance frameworks that evolve with technological progress. These technical and policy measures, in turn, are enabled by cross-disciplinary collaboration. Engineers are needed to translate legal and ethical requirements into technical specifications, while policymakers must understand the constraints and possibilities of technology to create effective regulations. Concurrently, ethicists ensure alignment with societal values, and human factors experts bridge the gap between technology and user trust by ensuring systems are usable and contextually adaptive. Ultimately, only by integrating these perspectives can we develop AI ecosystems that are both innovative and trustworthy, capable of navigating complex privacy risks throughout the entire AI lifecycle.

## 11.3 Vision for Co-Designed Privacy and Innovation for Human Benefit

Although the development of AI technologies and innovation may be shaped by privacy protections, research shows that enterprises adopting privacy-by-design principles benefit from lower compliance costs, stronger customer retention, and greater resilience to regulatory changes. In contrast, organizations that neglect privacy face higher risks and greater long-term expenses. For example, an IBM report highlights that ungoverned AI systems are not only more vulnerable to breaches but also significantly more costly when they are (Cost of a Data Breach 2025, IBM, n.d.). In the long term, embedding privacy into technological development builds trust, mitigates legal and ethical risks, and unlocks access to markets that would otherwise remain closed due to data sovereignty concerns. Rather than viewing privacy as a constraint, organizations are increasingly recognizing it as a keystone for resilient and inclusive innovation, which aligns technological progress with societal values and long-term sustainability.

**References**

Abadi, M., Chu, A., Goodfellow, I., McMahan, H. B., Mironov, I., Talwar, K., & Zhang, L. (2016). Deep learning with differential privacy. Proceedings of the 2016 ACM SIGSAC Conference on Computer and Communications Security, 308-318.

Acquisti, A., Brandimarte, L., & Loewenstein, G. (2015). Privacy and human behavior in the age of information. Science, 347(6221), 509-514.

Acquisti, A., Brandimarte, L., & Loewenstein, G. (2020). Secrets and likes: The drive for privacy and the difficulty of achieving it in the digital age. Journal of Consumer Psychology, 30(4), 736-758.

Aguirre, E., Roggeveen, A. L., Grewal, D., & Wetzels, M. (2016). The personalization-privacy paradox: implications for new media. Journal of consumer marketing, 33(2), 98-110.

Agrawal, G. (2024). Accountability, trust, and transparency in AI systems from the perspective of public policy: Elevating ethical standards. In AI healthcare applications and security, ethical, and legal considerations (pp. 148-162). IGI Global.

Akl, S. G., & Assem, I. (2020). Fully homomorphic encryption: a general framework and implementations. International Journal of Parallel, Emergent and Distributed Systems, 35(5), 493-498.

Alkhariji, L., Alhirabi, N., Alraja, M. N., Barhamgi, M., Rana, O., & Perera, C. (2021). Synthesising privacy by design knowledge toward explainable internet of things application designing in healthcare. ACM Transactions on Multimedia Computing, Communications, and Applications (TOMM), 17(2s), 1-29.

Allen, A. L. (1999). Privacy-as-data control: Conceptual, practical, and moral limits of the paradigm. Conn. L. Rev., 32, 861.

Alzoubi, Y. I., & Mishra, A. (2025). Differential privacy and artificial intelligence: potentials, challenges, and future avenues. EURASIP Journal on Information Security, 2025(1), 18.

Amershi, S., Weld, D., Vorvoreanu, M., Fourney, A., Nushi, B., Collisson, P., ... & Horvitz, E. (2019). Guidelines for human-AI interaction. In Proceedings of the 2019 chi conference on human factors in computing systems (pp. 1-13).

Asthana, S., Im, J., Chen, Z., & Banovic, N. (2024). " I know even if you don't tell me": Understanding Users' Privacy Preferences Regarding AI-based Inferences of Sensitive Information for Personalization. In Proceedings of the 2024 CHI Conference on Human Factors in Computing Systems (pp. 1-21).

Atadoga, A., Farayola, O. A., Ayinla, B. S., Amoo, O. O., Abrahams, T. O., & Osasona, F. (2024). A comparative review of data encryption methods in the USA and Europe. Computer Science & IT Research Journal, 5(2), 447-460.

Awad, N. F., & Krishnan, M. S. (2006). The personalization privacy paradox: an empirical evaluation of information transparency and the willingness to be profiled online for personalization. MIS quarterly, 13-28.

Aykhan Dadashov. (2023). Facial Recognition System is a Violation of Human Rights in the Context of the ECHR. Human Rights Brief, 27(1). https://digitalcommons.wcl.american.edu/hrbrief/vol27/iss1/7

Aziz, R., Banerjee, S., Bouzefrane, S., & Le Vinh, T. (2023). Exploring homomorphic encryption and differential privacy techniques towards secure federated learning paradigm. Future internet, 15(9), 310.

Baik, J. S. (2020). Data privacy against innovation or against discrimination?: The case of the California Consumer Privacy Act (CCPA). Telematics and Informatics, 52.

Bakare[1], S. S., Adeniyi, A. O., Akpuokwe, C. U., & Eneh[4], N. E. (2024). Data privacy laws and compliance: a comparative review of the EU GDPR and USA regulations.

Barocas, S., Crawford, K., Shapiro, A., & Wallach, H. (2017). The problem with bias: From allocative to representational harms in machine learning. In SIGCIS conference paper.

BBC News. (2017). Google's DeepMind makes AI application of NHS data a 'must-see'. https://www.bbc.com/news/technology-39301901

Bella, G., Librizzi, F., & Riccobene, S. (2008). A privacy paradigm that tradeoffs anonymity and trust. In 2008 16th International Conference on Software, Telecommunications and Computer Networks (pp. 384-388). IEEE.

Binns, R., Veale, M., Van Kleek, M., & Shadbolt, N. (2018). Like trainer, like bot? Inheritance of bias in algorithmic content moderation. In The Social, Cultural, and Ethical Dimensions of "Big Data" (pp. 1-12). Springer.

Bingley, W. J., Curtis, C., Lockey, S., Bialkowski, A., Gillespie, N., Haslam, S. A., ... & Worthy, P. (2023). Where is the human in human-centered AI? Insights from developer priorities and user experiences. Computers in Human Behavior, 141, 107617.

Binns, R., Van Kleek, M., Veale, M., Lyngs, U., Zhao, J., & Shadbolt, N. (2018). 'It's Reducing a Human Being to a Percentage' Perceptions of Justice in Algorithmic Decisions. In Proceedings of the 2018 Chi conference on human factors in computing systems (pp. 1-14).

Burton, S. L., Burrell, D., Nobles, C., White, Y. W., Dawson, M. E., Brown-Jackson, K. L., ... & Bessette, D. I. (2024). An in-depth qualitative interview: The impact of artificial intelligence (AI) on consent and transparency. In Multisector Insights in Healthcare, Social Sciences, Society, and Technology (pp. 248-269). IGI Global Scientific Publishing.

California Legislature. (2018). California Consumer Privacy Act of 2018, Cal. Civ. Code § 1798.100 et seq. https://leginfo.legislature.ca.gov/faces/codes_displayText.xhtml?lawCode=CIV&division=3.&title=1.81.5.&part=4.&chapter=&article=

California Legislature. (2020). California Privacy Rights Act of 2020, Cal. Civ. Code § 1798.100 et seq. https://leginfo.legislature.ca.gov/faces/codes_displayText.xhtml?division=3.&title=1.81.5.&part=4.&lawCode=CIV

Calo, R. (2012). Against Notice Skepticism in privacy and elsewhere. Notre Dame Law Review. vol. 87

Carlini, N., Tramer, F., Wallace, E., Jagielski, M., Herbert-Voss, A., Lee, K., ... & Raffel, C. (2021). Extracting training data from large language models. In 30th USENIX security symposium (USENIX Security 21) (pp. 2633-2650).

Carnegie Learning. (n.d.). MATHia software. https://www.carnegielearning.com/products/software-platform/mathia-learning-software/

Cath, C. (2018). Governing artificial intelligence: ethical, legal and technical opportunities and challenges. Philosophical Transactions of the Royal Society A: Mathematical, Physical and Engineering Sciences, 376(2133), 20180080.

Cheatham, B., Javanmardian, K., & Samandari, H. (2019). Confronting the risks of artificial intelligence. McKinsey Quarterly, 2(38), 1-9.

Chen, H. T., & Chen, W. (2015). Couldn't or wouldn't? The influence of privacy concerns and self-efficacy in privacy management on privacy protection. Cyberpsychology, Behavior, and Social Networking, 18(1), 13-19.

Chen, Y., Zhang, X., & Hu, L. (2024). A progressive prompt-based image-generative AI approach to promoting students' achievement and perceptions in learning ancient Chinese poetry. Educational Technology & Society, 27(2), 284-305.

Clemmensen, T. (2021). Human work interaction design for socio-technical theory and action. In Human Work Interaction Design: A Platform for Theory and Action (pp. 11-50). Cham: Springer International Publishing.

Cost of a data breach 2025 | IBM. (n.d.). https://www.ibm.com/reports/data-breach

Crawford, K. (2021). The atlas of AI: Power, politics, and the planetary costs of artificial intelligence. Yale University Press.

Dammu, P. P. S., Chalamala, S. R., & Singh, A. K. (2021). Explainable and Personalized Privacy Prediction. In CIKM Workshops.

Demianenko, V. (2019). Artificial intelligence systems in adaptive learning. Theory and practice of science education, 1(1).

Determann, L., & Tam, J. (2020). The California Privacy Rights Act of 2020: A broad and complex data processing regulation that applies to businesses worldwide. Journal of Data Protection & Privacy, 4(1), 7-21.

Distler, V., Lallemand, C., & Koenig, V. (2020). How acceptable is this? How user experience factors can broaden our understanding of the acceptance of privacy trade-offs. Computers in Human Behavior, 106, 106227.

Dodd, V., Police, V. D., & correspondent, crime. (2022). UK police use of live facial recognition unlawful and unethical, report finds. The Guardian. https://www.theguardian.com/technology/2022/oct/27/live-facial-recognition-police-study-uk

Dodiya, K., Radadia, S. K., & Parikh, D. (2024). Differential Privacy Techniques in Machine Learning for Enhanced Privacy Preservation.

Dwork, C., & Roth, A. (2014). The algorithmic foundations of differential privacy. Foundations and trends® in theoretical computer science, 9(3–4), 211-407.

Eiband, M., Buschek, D., Kremer, A., & Hussmann, H. (2019). The impact of placebic explanations on trust in intelligent systems. In Extended abstracts of the 2019 CHI conference on human factors in computing systems (pp. 1-6).

Erotokritou, S., Giannoulakis, I., Kafetzakis, E., & Kaltakis, K. (2024). Simplifying Differential Privacy for Non-Experts: The ENCRYPT Project Approach. In 2024 IEEE International Conference on Cyber Security and Resilience (CSR) (pp. 682-687). IEEE.

European Parliament and Council. (1995). Directive 95/46/EC of the European Parliament and of the Council of 24 October 1995 on the protection of individuals with regard to the processing of personal data and on the free movement of such data. Official Journal of the European Union, L 281, 31–50. http://data.europa.eu/eli/dir/1995/46/oj

European Parliament and Council. (2002). Directive 2002/58/EC of the European Parliament and of the Council of 12 July 2002 concerning the processing of personal data and the protection of privacy in the electronic communications sector (Directive on privacy and electronic communications). Official Journal of the European Union, L 201/37–47. http://data.europa.eu/eli/dir/2002/58/oj

European Union. (2016). Regulation (EU) 2016/679 of the European Parliament and of the Council of 27 April 2016 on the protection of natural persons with regard to the processing of personal data and on the free movement of such data, and repealing Directive 95/46/EC (General Data Protection Regulation). Official Journal of the European Union, L 119/1–88. http://data.europa.eu/eli/reg/2016/679/oj

European Union. (2024). Regulation (EU) 2024/1689 of the European Parliament and of the Council of 13 June 2024 laying down harmonised rules on artificial intelligence (Artificial Intelligence Act). Official Journal of the European Union, L, 1-89. https://eur-lex.europa.eu/legal-content/EN/TXT/?uri=OJ:L:2024:1689:TOC

Evans, D., Kolesnikov, V., & Rosulek, M. (2018). A Pragmatic Introduction to Secure Multi-Party Computation. Foundations and Trends in Privacy and Security, 2(2-3), 70-246.

Ezennaya-Gomez, S., Blumenthal, E., Eckardt, M., Krebs, J., Kuo, C., Porbeck, J., ... & Dittmann, J. (2022). Revisiting online privacy and security mechanisms applied in the in-app payment realm from the consumers' perspective. In Proceedings of the 17th International Conference on Availability, Reliability and Security (pp. 1-12).

Federal Ministry of Justice and Consumer Protection (Germany). (2023). Federal Data Protection Act (Bundesdatenschutzgesetz - BDSG). https://www.gesetze-im-internet.de/englisch_bdsg/

French National Assembly. (2018). Loi n° 78-17 du 6 janvier 1978 relative à l'informatique, aux fichiers et aux libertés [Law No. 78-17 of 6 January 1978 on Information Technology, Data Files and Civil Liberties]. Légifrance. https://www.legifrance.gouv.fr/loda/id/JORFTEXT000037085952/

Friedman, B., & Hendry, D. G. (2019). Value Sensitive Design: Shaping Technology with Moral Imagination. MIT Press.

Ghanem, S. M., & Moursy, I. A. (2019). Secure multiparty computation via homomorphic encryption library. In 2019 Ninth International Conference on Intelligent Computing and Information Systems (ICICIS) (pp. 227-232). IEEE.

Ghazi, B., Golowich, N., Kumar, R., Manurangsi, P., & Zhang, C. (2021). Deep learning with label differential privacy. Advances in neural information processing systems, 34, 27131-27145.

Groen, E. C., Feth, D., Polst, S., Tolsdorf, J., Wiefling, S., Iacono, L. L., & Schmitt, H. (2023). Achieving usable security and privacy through Human-Centered Design. In Human Factors in Privacy Research (pp. 83-113). Cham: Springer International Publishing.

Gudepu, B. K., & Eichler, R. (2024). The Role of AI in Enhancing Data Governance Strategies. International Journal of Acta Informatica, 3(1), 169-186.

Haakman, M., Cruz, L., Huijgens, H., & Van Deursen, A. (2021). AI lifecycle models need to be revised: An exploratory study in Fintech. Empirical Software Engineering, 26(5), 95.

Hagen, J., & Lysne, O. (2016). Protecting the digitized society—the challenge of balancing surveillance and privacy. The Cyber Defense Review, 1(1), 75-90.

Hayes, J., Melis, L., Danezis, G., & De Cristofaro, E. (2019). LOGAN: Evaluating privacy leakage of generative models using generative adversarial networks. Proceedings on Privacy Enhancing Technologies, 2019(1), 1-16.

Herington, J. (2020). Measuring fairness in an unfair world. In Proceedings of the AAAI/ACM Conference on AI, Ethics, and Society (pp. 286-292).

Hoffman, R. R., Mueller, S. T., Klein, G., & Litman, J. (2018). Metrics for explainable AI: Challenges and prospects. arXiv preprint arXiv:1812.04608.

Holland, S., Hosny, A., Newman, S., Joseph, J., & Chmielinski, K. (2020). The dataset nutrition label. Data protection and privacy, 12(12), 1.

Iachello, G., & Hong, J. (2007). End-user privacy in human–computer interaction. Foundations and Trends® in Human–Computer Interaction, 1(1), 1-137.

Italian Parliament. (2018). Testo coordinato del Codice in materia di protezione dei dati personali di cui al decreto legislativo 30 giugno 2003, n. 196 e successive modificazioni [Coordinated Text of the Legislative Decree No. 2003, n. 196: Personal Data Protection Code, and its subsequent amendments]. Gazzetta Ufficiale della Repubblica Italiana, 283. https://www.garanteprivacy.it/normativa/nazionale/testo-coordinato-del-codice-in-materia-di-protezione-dei-dati-personali-di-cui-al-decreto-legislativo-30-giugno-2003-n-196-e-sue-successive-modificazioni

Jacobsen, A., de Miranda Azevedo, R., Juty, N., Batista, D., Coles, S., Cornet, R., ... & Schultes, E. (2020). FAIR principles: interpretations and implementation considerations. Data intelligence, 2(1-2), 10-29.

Jain, P., Gyanchandani, M., & Khare, N. (2016). Big data privacy: a technological perspective and review. Journal of big data, 3(1), 25.

Jiang, B., Li, J., Yue, G., & Song, H. (2021). Differential privacy for industrial internet of things: Opportunities, applications, and challenges. IEEE Internet of Things Journal, 8(13), 10430-10451.

Jiang, Y., Zhou, Y., & Feng, T. (2022). A blockchain-based secure multi-party computation scheme with multi-key fully homomorphic proxy re-encryption. Information, 13(10), 481.

Jobin, A., Ienca, M., & Vayena, E. (2019). The global landscape of AI ethics guidelines. Nature machine intelligence, 1(9), 389-399.

Kamarinou, D., Millard, C., Singh, J., & Leenes, R. (2017). Machine learning with personal data. In Data protection and privacy: the age of intelligent machines (pp. 89-114). Oxford: Hart Publishing.

Kang, R., Dabbish, L., Fruchter, N., & Kiesler, S. (2015). {"My} data just goes {Everywhere:"} user mental models of the internet and implications for privacy and security. In Eleventh symposium on usable privacy and security (SOUPS 2015) (pp. 39-52).

Katz v. United States, 389 U.S. 347 (1967).

Khalid, N., Qayyum, A., Bilal, M., Al-Fuqaha, A., & Qadir, J. (2023). Privacy-preserving artificial intelligence in healthcare: Techniques and applications. Computers in Biology and Medicine, 158, 106848.

Khan, F., Khan, S., Tahir, S., Ahmad, J., Tahir, H., & Shah, S. A. (2021). Granular data access control with a patient-centric policy update for healthcare. Sensors, 21(10), 3556.

Kizilcec, R. F. (2016). How much information? Effects of transparency on trust in an algorithmic interface. In Proceedings of the 2016 CHI conference on human factors in computing systems (pp. 2390-2395).

Knijnenburg, B. P., Page, X., Wisniewski, P., Lipford, H. R., Proferes, N., & Romano, J. (2022). Modern socio-technical perspectives on privacy (p. 462). Springer Nature.

Kolesnyk, V., Molodetska, K., & Fedushko, S. (2025). From Connectivity to Security: Ensuring Cyber Resilience in Socio-technical Systems with Social Networking Services. In Developments in Information and Knowledge Management Systems for Business Applications: Volume 8 (pp. 121-147). Cham: Springer Nature Switzerland.

Kramer, M. W. (1999). Motivation to reduce uncertainty: A reconceptualization of uncertainty reduction theory. Management communication quarterly, 13(2), 305-316.

Krasnova, H., & Veltri, N. F. (2010). Privacy calculus on social networking sites: Explorative evidence from Germany and USA. In 2010 43rd Hawaii international conference on system sciences (pp. 1-10). IEEE.

Kumar, P. C., Zimmer, M., & Vitak, J. (2024). A roadmap for applying the contextual integrity framework in qualitative privacy research. Proceedings of the ACM on Human-Computer Interaction, 8(CSCW1), 1-29.

Kyi, L., Mhaidli, A., Santos, C. T., Roesner, F., & Biega, A. J. (2024, May). "It doesn't tell me anything about how my data is used": User Perceptions of Data Collection Purposes. In Proceedings of the 2024 CHI Conference on Human Factors in Computing Systems (pp. 1-12).

Lee, H. P., Yang, Y. J., Von Davier, T. S., Forlizzi, J., & Das, S. (2024). Deepfakes, phrenology, surveillance, and more! a taxonomy of ai privacy risks. In Proceedings of the 2024 CHI Conference on Human Factors in Computing Systems (pp. 1-19).

Lee, J. J., & Sayed, S. (2008). Culturally sensitive design. Design Connections-Knowledge, Value and Involvement through Design, 54-63.

Liao, Q. V., Gruen, D., & Miller, S. (2020). Questioning the AI: informing design practices for explainable AI user experiences. In Proceedings of the 2020 CHI conference on human factors in computing systems (pp. 1-15).

Liang, X., Xu, Y., Lin, Y. E., & Zhang, C. (2025). Federated split learning via dynamic aggregation and homomorphic encryption on non-IID data. The Journal of Supercomputing, 81(1), 63.

Liu, Z., Wang, W., Liang, H., & Yuan, Y. (2024). Enhancing data utility in personalized differential privacy: A fine-grained processing approach. In International Conference on Data Security and Privacy Protection (pp. 47-66). Singapore: Springer Nature Singapore.

Liu, Z., Zou, W., & Lin, C. (2025). Exploring the Influence of Privacy Concerns, AI Literacy, and Perceived Health Stigma on AI Chatbot Use in Healthcare: An Uncertainty Reduction Approach. Patient Education and Counseling, 109271.

Luger, E., & Rodden, T. (2013). An informed view on consent for UbiComp. In Proceedings of the 2013 ACM international joint conference on Pervasive and ubiquitous computing (pp. 529-538).

Malik, W., Gul, S., & Qureshi, G. M. (2025). Regulating Artificial Intelligence: Challenges for Data Protection and Privacy in Developing Nations. Journal of Social Signs Review, 3(05), 95-108.

Mammen, P. M. (2021). Federated learning: Opportunities and challenges. arXiv preprint arXiv:2101.05428.

Mantelero, A., & Esposito, M. S. (2021). An evidence-based methodology for human rights impact assessment in the development of AI data-intensive systems. Computer Law & Security Review, 41, 105561.

McAran, D. (2021). Privacy, ethics, trust, and UX challenges as reflected in google's people and AI guidebook. In International Conference on Human-Computer Interaction (pp. 588-599). Cham: Springer International Publishing.

Mehrabi, N., Morstatter, F., Saxena, N., Lerman, K., & Galstyan, A. (2021). A survey on bias and fairness in machine learning. ACM computing surveys (CSUR), 54(6), 1-35.

Meo, S. A., & Talha, M. (2019). Turnitin: Is it a text matching or plagiarism detection tool?. Saudi journal of anaesthesia, 13(Suppl 1), S48-S51.

Mohseni, S., Zarei, N., & Ragan, E. D. (2021). A multidisciplinary survey and framework for design and evaluation of explainable AI systems. ACM Transactions on Interactive Intelligent Systems (TiiS), 11(3-4), 1-45.

Monteleone, S., van Bavel, R., Rodríguez-Priego, N., & Esposito, G. (2015). Nudges to privacy behaviour: Exploring an alternative approach to privacy notices. JRC Science and Policy Report. Luxembourg, Luxembourg: Publications Office of the European Union.

Morehouse, K. N., Kurdi, B., & Nosek, B. A. (2024). Responsible data sharing: Identifying and remedying possible re-identification of human participants. American Psychologist.

Naehrig, M., Lauter, K., & Vaikuntanathan, V. (2011). Can Homomorphic Encryption be Practical? Proceedings of the 3rd ACM Workshop on Cloud Computing Security Workshop (pp. 113-124).

National Institute of Standards and Technology. (2020). Privacy framework: A tool for improving privacy through enterprise risk management (Version 1.0). U.S. Department of Commerce. https://www.nist.gov/privacy-framework/privacy-framework

National People's Congress of the People's Republic of China. (2021). Personal information protection law of the People's Republic of China. Standing Committee of the NPC. http://www.npc.gov.cn/npc/c2/c30834/202108/t20210820_313090.html

Navarro, C. L. A., Damen, J. A., Takada, T., Nijman, S. W., Dhiman, P., Ma, J., ... & Hooft, L. (2021). Risk of bias in studies on prediction models developed using supervised machine learning techniques: systematic review. bmj, 375.

Nepal, S., Ko, R. K., Grobler, M., & Camp, L. J. (2022). Human-centric security and privacy. Frontiers in big Data, 5, 848058.

News. (2020). Proctorio faces backlash after CEO breaches privacy of UBC student. The Peak. https://the-peak.ca/2020/07/proctorio-faces-backlash-after-ceo-breaches-privacy-of-ubc-student/

Okpo, O., & Joseph, A. S. (2025). ARTIFICIAL INTELLIGENCE AND ETHICAL BOUNDARIES: A DEONTOLOGICAL CRITIQUE OF DATA PRIVACY VIOLATIONS. SAPIENTIA, 21(1), 124.

Olatunji, I. E., Rauch, J., Katzensteiner, M., & Khosla, M. (2024). A review of anonymization for healthcare data. Big data, 12(6), 538-555.

Oliveira, M., Yang, J., Griffiths, D., Bonnay, D., & Kulshrestha, J. (2025). Browsing behavior exposes identities on the Web. Scientific Reports, 15(1), 36066.

Owen, A., & Mattews, A. (2024). Privacy Rights and Interoperability: A Comparative Study of Regulations.

Pan, K., Ong, Y. S., Gong, M., Li, H., Qin, A. K., & Gao, Y. (2024). Differential privacy in deep learning: A literature survey. Neurocomputing, 589, 127663.

Polasik, M., Widawski, P., & Lis, A. (2022). Challenger bank as a new digital form of providing financial services to retail customers in the EU internal market: The case of Revolut. In The Digitalization of Financial Markets. Taylor & Francis.

Raab, C., & Wright, D. (2012). Surveillance: extending the limits of privacy impact assessment. In Privacy impact assessment (pp. 363-383). Dordrecht: Springer Netherlands.

Razaque, A., Hariri, S., & Yoo, J. (2025). AI-Driven User Interface Design: Enhancing Digital Learning and Skill Development. Available at SSRN 5114814.

Rees, L., Safi, R., & Lim, S. L. (2022). How much will you share?: Exploring attitudinal and behavioral nudges in online private information sharing. Journal of Experimental Psychology: Applied, 28(4), 775.

Ribeiro, M. T., Singh, S., & Guestrin, C. (2016). " Why should i trust you?" Explaining the predictions of any classifier. In Proceedings of the 22nd ACM SIGKDD international conference on knowledge discovery and data mining (pp. 1135-1144).

Riedl, M. O. (2019). Human‐centered artificial intelligence and machine learning. Human behavior and emerging technologies, 1(1), 33-36.

Rigaki, M., & Garcia, S. (2023). A survey of privacy attacks in machine learning. ACM Computing Surveys, 56(4), 1-34.

Ropek, L. (2023). New York City schools ban ChatGPT to head off a cheating epidemic. Gizmodo. Disponible en: http://bit.ly/3kp8Ha9.

Rudolph, M., Feth, D., & Polst, S. (2018). Why users ignore privacy policies–a survey and intention model for explaining user privacy behavior. In International Conference on Human-Computer Interaction (pp. 587-598). Cham: Springer International Publishing.

Sadek, M., Constantinides, M., Quercia, D., & Mougenot, C. (2024). Guidelines for integrating value sensitive design in responsible AI toolkits. In Proceedings of the 2024 CHI Conference on Human Factors in Computing Systems (pp. 1-20).

Schoenherr, J. R., Abbas, R., Michael, K., Rivas, P., & Anderson, T. D. (2023). Designing AI using a human-centered approach: Explainability and accuracy toward trustworthiness. IEEE Transactions on Technology and Society, 4(1), 9-23.

Schomakers, E. M., Lidynia, C., & Ziefle, M. (2022). The role of privacy in the acceptance of smart technologies: Applying the privacy calculus to technology acceptance. International Journal of Human–Computer Interaction, 38(13), 1276-1289.

Schoormann, T., Möller, F., Chandra Kruse, L., & Otto, B. (2024). BAUSTEIN—A design tool for configuring and representing design research. Information Systems Journal, 34(6), 1871-1901.

Shastri, S., Banakar, V., Wasserman, M., Kumar, A., & Chidambaram, V. (2019). Understanding and benchmarking the impact of GDPR on database systems. arXiv preprint arXiv:1910.00728.
39

Shneiderman, B. (2020). Human-centered artificial intelligence: Three fresh ideas. AIS Transactions on Human-Computer Interaction, 12(3), 109-124.

Shneiderman, B. (2022). Human-Centered AI. Oxford University Press.

Shokri, R., Stronati, M., Song, C., & Shmatikov, V. (2017). Membership inference attacks against machine learning models. In 2017 IEEE symposium on security and privacy (SP) (pp. 3-18). IEEE.

Simonsen, J., & Robertson, T. (Eds.). (2013). Routledge international handbook of participatory design (Vol. 711). New York: Routledge.

Smit, K., Zoet, M., & van Meerten, J. (2020). A review of AI principles in practice.

Smith, H. J., Milberg, S. J., & Burke, S. J. (1996). Information privacy: Measuring individuals' concerns about organizational practices. MIS quarterly, 20(2), 167-196.

Soeder, M. O. (2019). Privacy challenges and approaches to the consent dilemma. Available at SSRN 3442612.

Soh, S. Y. (2019). Privacy nudges: An alternative regulatory mechanism to informed consent for online data protection behaviour. Eur. Data Prot. L. Rev., 5, 65.

Solove, D. J. (2006). A taxonomy of privacy. University of Pennsylvania Law Review, 154(3), 477-560.

Springer, A., & Whittaker, S. (2018). Progressive disclosure: designing for effective transparency. arXiv preprint arXiv:1811.02164.

Staab, R., Vero, M., Balunović, M., & Vechev, M. (2023). Beyond memorization: Violating privacy via inference with large language models. arXiv preprint arXiv:2310.07298.

Sun, L., & Yang, B. (2021). Your privacy preference matters: a qualitative study envisioned for homecare. In 2021 IEEE Symposium on Computers and Communications (ISCC) (pp. 1-7). IEEE.

Sun, L., Yang, B., Kindt, E., & Chu, J. (2024). Privacy barriers in health monitoring: scoping review. JMIR nursing, 7, e53592.

Suresh, H., & Guttag, J. (2021). A framework for understanding sources of harm throughout the machine learning life cycle. In Proceedings of the 1st ACM Conference on Equity and Access in Algorithms, Mechanisms, and Optimization (pp. 1-9).

Tavani, H. T. (2007). Philosophical theories of privacy: Implications for an adequate online privacy policy. Metaphilosophy, 38(1), 1-22.

Teltzrow, M., & Kobsa, A. (2004). Impacts of user privacy preferences on personalized systems: a comparative study. In Designing personalized user experiences in eCommerce (pp. 315-332). Dordrecht: Springer Netherlands.

Usmani, U. A., Happonen, A., & Watada, J. (2023). Human-centered artificial intelligence: Designing for user empowerment and ethical considerations. In 2023 5th international congress on human-computer interaction, optimization and robotic applications (HORA). IEEE.

Veale, M., Binns, R., & Edwards, L. (2018). Algorithms that remember: Model inversion attacks and data protection law. Philosophical Transactions of the Royal Society A: Mathematical, Physical and Engineering Sciences, 376 (2133), 20180083.

Wachter, S., Mittelstadt, B., & Floridi, L. (2017). Why a right to explanation of automated decision-making does not exist in the general data protection regulation. International data privacy law, 7(2), 76-99.

Wacnik, P., Daly, S. R., & Verma, A. (2025). Participatory design: a systematic review and insights for future practice. Design Science, 11, e21.

Wang, B., & Gong, N. Z. (2018). Stealing hyperparameters in machine learning. In 2018 IEEE symposium on security and privacy (SP) (pp. 36-52). IEEE.

Wang, K., Zhang, G., Zhou, Z., Wu, J., Yu, M., Zhao, S., ... & Liu, Y. (2025). A comprehensive survey in llm (-agent) full stack safety: Data, training and deployment. arXiv preprint arXiv:2504.15585.

Wei, K., Li, J., Ma, C., Ding, M., Chen, W., Wu, J., ... & Poor, H. V. (2023). Personalized federated learning with differential privacy and convergence guarantee. IEEE Transactions on Information Forensics and Security, 18, 4488-4503.

Weidinger, L., Mellor, J., Rauh, M., Griffin, C., Uesato, J., Huang, P. S., ... & Gabriel, I. (2021). Ethical and social risks of harm from language models. arXiv preprint arXiv:2112.04359.

Wongso, B., Lienaka, K. N., Firstian, V., & Magdalena, Y. (2024). User-centered design in AI applications: a systematic literature review. In 2024 International Conference on Information Management and Technology (ICIMTech) (pp. 524-529). IEEE.

Xu, W. (2019). Toward human-centered AI: a perspective from human-computer interaction. interactions, 26(4), 42-46.

Xu, W., Gao, Z., & Dainoff, M. (2023). An HCAI methodological framework: Putting it into action to enable human-centered AI. arXiv preprint arXiv:2311.16027.

Xu Wei. (2024).User-Centered Design（IX）：A "User Experience 3.0" Paradigm Framework in the Intelligence Era. Chinese Journal of Applied Psychology, 30(2), 99-117.

Xu, W., Gao, Z., Ge, L. (2024). New research paradigms and agenda of human factors science in the intelligence era. Acta Psychologica Sinica, 56(3), 363-382.

Yang, Y., Zhang, B., Guo, D., Du, H., Xiong, Z., Niyato, D., & Han, Z. (2024). Generative AI for secure and privacy-preserving mobile crowdsensing. IEEE Wireless Communications, 31(6), 29-38.

Yeung, K. (2020). Recommendation of the council on artificial intelligence (OECD). International legal materials, 59(1), 27-34.

Young, I. (2008). Mental models: aligning design strategy with human behavior. Rosenfeld Media.

Zaeem, R. N., & Barber, K. S. (2020). The effect of the GDPR on privacy policies: Recent progress and future promise. ACM Transactions on Management Information Systems, 12(1), 1-20.

Zhong, B., Sun, T., Zhou, Y., & Xie, L. (2024). Privacy matters: reexamining internet privacy concern among social media users in a cross-cultural setting. Atlantic Journal of Communication, 32(2), 180-197.

Zhu, L., Liu, Z., & Han, S. (2019). Deep Leakage from Gradients. Advances in Neural Information Processing Systems, 32 (NeurIPS).

Zuboff, S. (2019). The age of surveillance capitalism. PublicAffairs.